\documentclass[a4]{JHEP3}
\usepackage{amsmath}
\usepackage{feynmf}

\newcommand{\cl}[1]{#1}
\newcommand{\al}[1]{\bar{#1}}

\newcommand{\ci}[1]{#1}
\newcommand{\ai}[1]{\dot{#1}}

\newcommand{\cd}{\cl{\text{d}}_{\ci{+}}}
\newcommand{\ad}{\al{\text{d}}_{\ai{+}}}

\newcommand{\pdd}[2]{\partial_{\ci{#1}\ai{#2}}}
\newcommand{\dpp}{\pdd{+}{+}}
\newcommand{\dpm}{\pdd{+}{-}}
\newcommand{\dmp}{\pdd{-}{+}}
\newcommand{\dmm}{\pdd{-}{-}}

\newcommand{\dbox}{\Box}

\newcommand{\mdd}[3]{{#1}_{\ci{#2}\ai{#3}}}
\newcommand{\mpp}[1]{\mdd{#1}{+}{+}}
\newcommand{\mpm}[1]{\mdd{#1}{+}{-}}
\newcommand{\mmp}[1]{\mdd{#1}{-}{+}}

\newcommand{\ctheta}{\cl{\theta}}
\newcommand{\atheta}{\al{\theta}}

\newcommand{\cthetau}[1]{\ctheta^{\ci{#1}}}
\newcommand{\athetau}[1]{\atheta^{\ai{#1}}}

\newcommand{\cZ}{\cl{Z}}
\newcommand{\aZ}{\al{Z}}
\newcommand{\cpsi}{\cl{\psi}_{\ci{-}}}
\newcommand{\apsi}{\al{\psi}_{\ai{-}}}
\newcommand{\cJ}{\cl{J}_{\ci{-}}}
\newcommand{\aJ}{\al{J}_{\ai{-}}}
\newcommand{\ceta}{\cl{\eta}}
\newcommand{\aeta}{\al{\eta}}

\newcommand{\vphi}{\phi}
\newcommand{\vJ}{\mathcal{J}}

\newcommand{\di}{\text{d}}

\newcommand{\gp}[1]{\widetilde{#1}}

\newcommand{\ndd}[2]{\mdd{n}{#1}{#2}}

\newcommand{\muu}[3]{{#1}^{\ci{#2}\ai{#3}}}

\newcommand{\nuu}[2]{\muu{n}{#1}{#2}}

\newcommand{\adu}[3]{{{#1}_{\ai{#2}}}^{\ai{#3}}}

\newcommand{\cdu}[3]{{{#1}_{\ci{#2}}}^{\ci{#3}}}

\newcommand{\add}[3]{{\al{#1}_{\ai{#2}\ai{#3}}}}

\newcommand{\cdd}[3]{{\cl{#1}_{\ci{#2}\ci{#3}}}}

\title{SIM(2) and supergraphs}

\author{Stanislav Petr\'a\v{s}, Rikard von Unge and Ji\v{r}\'{\i} Voh\'anka\\
Masaryk University, Institute for Theoretical Physics\\
Kotl\'a\v{r}sk\'a 2, 611 37 Brno, Czech Republic\\
Email: \email{standa@physics.muni.cz},\email{unge@physics.muni.cz},
\email{vohanka@physics.muni.cz}
}

\abstract{We construct Feynman rules and Supergraphs in SIM(2) superspace. To test
our methods we
perform a one-loop calculation of the effective action of the SIM(2) supersymmetric
Wess-Zumino model including a term which explicitly breaks Lorentz invariance.
The renormalization of the model is also discussed.}


\preprint{}

\begin{document}
\begin{fmffile}{feynmf_base}

\section{Introduction}
Recently Cohen and Glashow observed \cite{cg,cg1} that the assumption that physics
is invariant
under the SIM(2) subgroup of the Lorentz group is entirely compatible with
experiments. This opens up new possibilities in particle phenomenology, in
particular concerning neutrino masses.

Following the initial papers, further progress was made in \cite{Gibbons} where 
the question of gravity in SIM(2) symmetric theories was addressed. Also, the
connection to noncommutative geometry was discussed in \cite{Sheikh}. Recently, it
was claimed \cite{Das} that the setup of Cohen and Glashow gives predictions for
the Thomas precession which are incompatible with experiments.

The possibility of SIM(2) supersymmetric theories was considered in \cite{cf} and
subsequently given a superspace formulation in \cite{uam}. In this paper we
use the superspace formulation to develop Feynman rules for SIM-supersymmetric
theories. To test our formalism we perform an explicit calculation of the
one loop effective action including the renormalization for the Wess-Zumino
model with an added Lorentz breaking but SIM-supersymmetric term.
We find that the physics of this model is not
dramatically different from the physics of the Wess-Zumino model. The
model is infrared free with the presence of a Landau pole. Furthermore, in
the limit where the Lorentz breaking goes to zero, we recover the standard
results from the Wess-Zumino model. Thus we conclude that the Feynman
rules that we propose are consistent. We also find that the Lorentz symmetry
breaking mass term renormalize to zero more slowly than the Lorentz
symmetry preserving mass term of the Wess-Zumino model.

This paper is organized as follows. In section \ref{setup} we introduce
SIM-superspace, in section \ref{sec:genfunc} we present the action that we will use
in our calculations, in section \ref{sec:fr} we present the Feynman rules of our theory.
In section \ref{sec:d} we show how to perform many of the calculations directly on
the supergraphs, so called d-algebra manipulations. As a direct consequence of our
Feynman rules we derive a perturbative nonrenormalization theorem and discuss its
consequences in section \ref{sec:nonren}. In section \ref{sec:morex} we compute
a particular Feynman diagram in detail. In section \ref{sec:all} we present the result
for the effective action and in section \ref{sec:ren} we discuss renormalization
of the model. Finally we conclude in section \ref{sec:concl}.
\section{The basic setup}
\label{setup}
The SIM(2) group is the subgroup of the four dimensional Lorentz group that preserves a fixed null fourvector $n$ up to rescaling. A summary of all relevant facts can be found for example in \cite{cg}.

The SIM(2) group is a solvable group so its irreducible representations are one dimensional. The conditions $\ndd{\alpha}{\alpha}\cl{\epsilon}^{\ci{\alpha}} = \ndd{\alpha}{\alpha}\al{\epsilon}^{\ai{\alpha}} = 0$ single out the invariant subspaces in the space of the left and right handed Weyl spinors.

To solve the above conditions it is convenient to choose a second fourvector $\tilde{n}$ that obeys $n\cdot\tilde{n}=1$ and define the set of orthogonal projectors
\begin{align*}
&{{P_{1}}_{\alpha}}^{\gamma}=\ndd{\alpha}{\beta}\muu{\tilde{n}}{\gamma}{\beta}, &\adu{\overline{P}_{1}}{\alpha}{\gamma}=\ndd{\beta}{\alpha}\muu{\tilde{n}}{\beta}{\gamma},\nonumber \\
&{{P_{2}}_{\alpha}}^{\gamma}=\mdd{\tilde{n}}{\alpha}{\beta}\nuu{\gamma}{\beta}, &\adu{\overline{P}_{2}}{\alpha}{\gamma}=\mdd{\tilde{n}}{\beta}{\alpha}\nuu{\beta}{\gamma},
\end{align*}
which satisfy the relation ${P_{1\alpha}}^{\beta} + {P_{2\alpha}}^{\beta} = \delta_\alpha^\beta$. The properties of the projectors can be verified using the identity
\begin{equation*}
\muu{n_{1}}{\alpha}{\beta}\mdd{n_{2}}{\gamma}{\beta}+\muu{n_{2}}{\alpha}{\beta}\mdd{n_{1}}{\gamma}{\beta}=(n_{1}\cdot n_{2}){\delta^{\alpha}}_{\gamma}.
\end{equation*}
These projectors can then be used to decompose any spinor $\theta = P_{1}\theta + P_{2}\theta = \varepsilon + \varepsilon'$
and the projection $\epsilon = P_1\theta$ then provides a solution of the condition $\ndd{\alpha}{\alpha}\cl{\epsilon}^{\ci{\alpha}} = 0$. Similar relations hold also for the conjugated projectors.

The SIM(2) superalgebra consists of the generators of the SIM(2) rotations $\cdd{J}{\alpha}{\beta}$, $\add{J}{\alpha}{\beta}$ as well as translations $\mdd{P}{\alpha}{\alpha}$ and supertranslations
\begin{align}
S_{\alpha}&=\cdu{P_{2}}{\alpha}{\gamma}Q_{\gamma}, &\al{S}_{\ai{\alpha}}&= \adu{\overline{P}_{2}}{\alpha}{\gamma}\al{Q}_{\ai{\gamma}},
\end{align}
defined as projections of the generators of Poincare supertranslations. 
They satisfy the relations
\begin{align}
\lbrace S_{\alpha}, \al{S}_{\ai{\alpha}} \rbrace &= (-\dfrac{1}{2}\tilde{n}^{2}\ndd{\alpha}{\alpha}+\mdd{\tilde{n}}{\alpha}{\alpha})\sqrt{2}(n\cdot P),\nonumber \\
\lbrace S_{\alpha}, S_{\beta} \rbrace &= \lbrace \al{S}_{\ai{\alpha}}, \al{S}_{\ai{\beta}} \rbrace = 0, \nonumber \\
\end{align}
which are consequences of the relations $\lbrace Q_{\alpha}, \al{Q}_{\ai{\beta}}\rbrace = P_{\alpha\ai{\beta}}$, $\lbrace Q_{\alpha}, Q_{\beta}\rbrace = \lbrace \al{Q}_{\ai{\alpha}}, \al{Q}_{\ai{\beta}}\rbrace = 0$.

We also define the projections of the $N=1$ super covariant derivatives $D_{\alpha}$ and $\al{D}_{\ai{\alpha}}$
\begin{align}
d'_{\alpha} &= \cdu{P_{2}}{\alpha}{\gamma}D_{\gamma}, & 
\al{d}'_{\ai{\alpha}}&= \adu{\overline{P}_{2}}{\alpha}{\gamma}\al{D}_{\ai{\gamma}},\\  
q'_{\alpha} &= \cdu{P_{1}}{\alpha}{\gamma}D_{\gamma}, &
\al{q}'_{\ai{\alpha}}&= \adu{\overline{P}_{1}}{\alpha}{\gamma}\al{D}_{\ai{\gamma}},
\end{align}
Such projections of the covariant derivatives then fulfill the following algebra
\begin{align}
&\lbrace d'_{\alpha}, \al{d}'_{\ai{\alpha}} \rbrace = i(-\dfrac{1}{2}\tilde{n}^{2}\ndd{\alpha}{\alpha}+\mdd{\tilde{n}}{\alpha}{\alpha})\sqrt{2}(n\cdot\partial),\nonumber \\
&\lbrace d'_{\alpha}, d'_{\beta} \rbrace = 0,\nonumber \\
&\lbrace d'_{\alpha}, q'_{\beta} \rbrace = 0,\nonumber \\
&\lbrace d'_{\alpha}, \al{q}'_{\ai{\alpha}} \rbrace = i(\dfrac{1}{\sqrt{2}}\tilde{n}^{2}\ndd{\alpha}{\alpha}(n\cdot\partial) - \mdd{\tilde{n}}{\alpha}{\beta}\ndd{\lambda}{\alpha}\partial^{\lambda\ai{\beta}}),\nonumber \\
&\lbrace q'_{\alpha}, \al{q}'_{\ai{\alpha}} \rbrace = i(\sqrt{2}\ndd{\alpha}{\alpha}(\tilde{n}\cdot \partial) - \dfrac{1}{\sqrt{2}}\tilde{n}^{2}\ndd{\alpha}{\alpha}(n\cdot\partial)),\nonumber \\
&\lbrace q'_{\alpha}, q'_{\beta} \rbrace = 0.
\end{align}

Because the SIM(2) superalgebra has only half of the supertranslations of the super-Poincare superalgebra it is natural to represent it on a superspace having only half of the Grassman coordinates. We can define the relevant Grassman variables by requiring that they satisfy
$\ndd{\alpha}{\alpha}\cl{\theta}^{\ci{\alpha}} = \ndd{\alpha}{\alpha}\al{\theta}^{\ai{\alpha}} = 0$. It is useful to define the projected Grassman coordinates
\begin{align*}
\cl{\varepsilon}_{\ci{\alpha}} &= \cdu{P_1}{\alpha}{\beta}\cl{\theta}_{\ci{\beta}}, &
\cl{\varepsilon'}_{\ci{\alpha}} &= \cdu{P_2}{\alpha}{\beta}\cl{\theta}_{\ci{\beta}}, \\
\al{\varepsilon}_{\ai{\alpha}} &= \adu{\overline{P}_1}{\alpha}{\beta}\al{\theta}_{\ai{\beta}}, &
\al{\varepsilon'}_{\ai{\alpha}} &= \adu{\overline{P}_2}{\alpha}{\beta}\al{\theta}_{\ai{\beta}}.
\end{align*}
Then any super-Poincare superfield gives rise to SIM(2) superfields by setting $\cl{\varepsilon'}$ and $\al{\varepsilon'}$ to zero.

For example, let $\Phi$ ($\bar{\Phi}$) be an ordinary (anti)chiral scalar superfields. The SIM(2) projection of such fields, depending only on the $\cl{\varepsilon}$ and $\al{\varepsilon}$ spinors, is defined as
\begin{align}
\cZ(x,\varepsilon,\bar{\varepsilon}) &= \Phi(x,\varepsilon + \varepsilon', \bar{\varepsilon} + \bar{\varepsilon}')|_{\varepsilon',\bar{\varepsilon}'},
\nonumber \\
\aZ(x,\varepsilon,\bar{\varepsilon}) &= \bar{\Phi}(x,\varepsilon + \varepsilon', \bar{\varepsilon} + \bar{\varepsilon}')|_{\varepsilon',\bar{\varepsilon}'},
\nonumber\\
\psi_\alpha(x,\varepsilon,\bar{\varepsilon}) &= q'_\alpha \Phi(x,\varepsilon + \varepsilon', \bar{\varepsilon} + \bar{\varepsilon}')|_{\varepsilon',\bar{\varepsilon}'},
\\
\bar{\psi}_{\dot{\alpha}}(x,\varepsilon,\bar{\varepsilon}) &= \bar{q}'_{\dot{\alpha}} \bar\Phi(x,\varepsilon + \varepsilon', \bar{\varepsilon} + \bar{\varepsilon}')|_{\varepsilon',\bar{\varepsilon}'},\nonumber
\end{align}
where $|_{\varepsilon',\bar{\varepsilon}'}$ means that we put $\cl{\varepsilon}' = \al{\varepsilon}' = 0$.
The properties of the algebra of covariant derivatives, the chirality condition $\bar{D}_{\ai{\alpha}}\Phi=0$ of the superfield $\Phi$ and antichirality condition $D_{\alpha}\bar{\Phi}=0$ of the superfield $\bar{\Phi}$ then give us the following conditions for the SIM superfields $\cZ$, $\aZ$, $\psi_{\alpha}$ and $\bar{\psi}_{\ai{\alpha}}$:
\begin{align}
\al{d}_{\ai{\alpha}}\cZ &= 0, &
\cl{d}_{\ci{\alpha}}\aZ &= 0, \nonumber \\
\al{d}_{\ai{\alpha}}\cl{\psi}_{\ci{\alpha}}&= \lbrace \al{d}'_{\ai{\alpha}}, \cl{q}'_{\ci{\alpha}} \rbrace \cZ, &
\cl{d}_{\ci{\alpha}}\al{\psi}_{\ai{\alpha}}&= \lbrace \cl{d}'_{\ci{\alpha}}, \al{q}'_{\ai{\alpha}} \rbrace \aZ,
\end{align}
where $\cl{d}_{\ci{\alpha}} = \cl{d}'_{\ci{\alpha}}|_{\cl{\varepsilon}',\al{\varepsilon}'}$ and $\al{d}_{\ai{\alpha}} = \al{d}'_{\ai{\alpha}}|_{\cl{\varepsilon}',\al{\varepsilon}'}$ are the covariant derivatives acting in the SIM(2) superspace. They satisfy the same relation as $\cl{d}'_{\ci{\alpha}}$, $\al{d}'_{\ai{\alpha}}$
\begin{align*}
 \{\cl{d}_{\ci{\alpha}}, \al{d}_{\ai{\alpha}}\} &= i(-\dfrac{1}{2}\tilde{n}^{2}\ndd{\alpha}{\alpha}+\mdd{\tilde{n}}{\alpha}{\alpha})\sqrt{2}(n\cdot\partial), &
 \{\cl{d}_{\ci{\alpha}}, \cl{d}_{\ci{\beta}}\} &= \{\al{d}_{\ai{\alpha}}, \al{d}_{\ai{\beta}}\} = 0.
\end{align*}
We may choose coordinates so that the lightlike fourvector $n^{\alpha\dot\alpha}$ becomes $n^{\ci{+}\ai{+}} = 1, n^{\ci{+}\ai{-}} = n^{\ci{-}\ai{+}} = n^{\ci{-}\ai{-}} = 0$. The second fourvector $\tilde{n}_{\alpha\dot\alpha}$ can then be chosen to be $\tilde{n}_{\ci{+}\ai{+}} = 1, \tilde{n}_{\ci{+}\ai{-}} = \tilde{n}_{\ci{-}\ai{+}} = \tilde{n}_{\ci{-}\ai{-}} = 0$.
After this simplification, which we will use from now on, we get:
\begin{align}
\lbrace d'_{+}, \al{d}'_{\ai{+}} \rbrace &= i\partial_{+\ai{+}},\nonumber\\
\lbrace q'_{-}, \al{q}'_{\ai{-}} \rbrace &= i\partial_{-\ai{-}},\\
\lbrace d'_{+}, \al{q}'_{\ai{-}} \rbrace &= i\partial_{+\ai{-}}, \nonumber
\end{align}
as well as
\begin{align}
\ad\cpsi' &= i\dmp\cZ, &
\cd\apsi' &= i\dpm\aZ, \nonumber \\
\end{align}

Because of the constraint on the $\cpsi'$ superfield, it does not transform irreducibly under SIM(2) rotations but rather it mixes with the
$\cd\cZ$ component of the $\cZ$ SIM-superfield.
It is therefore useful to replace the SIM-superfileds $\cpsi'$, $\apsi'$ with the new superfields $\cpsi$, $\apsi$ related to the original ones as
\begin{align}
\cpsi = \cpsi' - \dfrac{\dmp}{\dpp}\cd\cZ,\nonumber \\
\apsi = \apsi' - \dfrac{\dpm}{\dpp}\ad\aZ.
\end{align} 
It is straightforward to verify that the new superfield $\cpsi$ and $\apsi$ transform into themselves under SIM(2) rotations. Furthermore they are {\em chiral} as SIM-superfields
\begin{align*}
 \ad\cpsi &= 0, &
 \cd\apsi &= 0.
\end{align*}

We may now rewrite the well known Wess-Zumino action for a chiral scalar superfield in the SIM(2) superspace formalism. The Wess-Zumino action is
\begin{equation}
 \begin{split}
S&=\int d^{4}xd^{2}\theta d^{2}\al{\theta}\Phi\al{\Phi} + \dfrac{1}{2}M\int d^{4}xd^{2}\theta\Phi^{2} +\dfrac{1}{2}M\int d^{4}xd^{2}\al{\theta} \al{\Phi}^{2}+ \\ &+ \dfrac{\lambda}{3!}\int d^{4}xd^{2}\theta\Phi^{3}+ \dfrac{\lambda}{3!}M\int d^{4}xd^{2}\al{\theta} \al{\Phi}^{3}.
 \end{split}
\end{equation}

We illustrate the procedure on the first term in the action. The idea is to rewrite the Grassman measure $\cl{D}^2\al{D}^2$ in terms of $\cd'$, $\ad'$, $\cl{q}'_{\ci{-}}$, $\al{q}'_{\ai{-}}$ and then perform the integration over $\cl{q}'_{\ci{-}}$, $\al{q}'_{\ai{-}}$
\begin{equation}
 \begin{split}
\int d^{4}xd^{2}\theta d^{2}\al{\theta}\Phi\al{\Phi} &= \int d^{4}xD^{2}\al{D}^{2}\Phi\al{\Phi}|_{\theta\al{\theta}} =\int d^{4}x i^{2}\cd'\cl{q}'_{-}\ad'\al{q}'_{\ci{-}}(\cl{\Phi}\al{\Phi})|_{\cl{\varepsilon}\al{\varepsilon}\cl{\varepsilon}'\al{\varepsilon}'}=\\
&= \int d^{4}x\cd'\ad'\left.\left[ (\cl{q}'_{\ci{-}}\cl{\Phi})(\al{q}'_{\ai{-}}\al{\Phi}) +\cl{\Phi}i\dmm\al{\Phi}\right]\right|_{\cl{\varepsilon}\al{\varepsilon}\cl{\varepsilon}'\al{\varepsilon}'}= \\
&= \int d^{4}x\cd\ad\left[ \cpsi\apsi + \cZ\frac{\dbox}{i\dpp}\aZ \right].
 \end{split}
\end{equation}
The same procedure is used to rewrite the other terms in the Wess-Zumino action in the SIM(2) superspace formalism.

Thus, the Wess-Zumino model in SIM(2) superspace formalism has two chiral superfields $\cZ$, $\cpsi$ and two antichiral superfields $\aZ$, $\apsi$. 
The superfields $\cZ$, $\aZ$ are Grassman even, while the superfields $\cpsi$, $\apsi$ are Grassman odd.

The action $S$ can be split into two parts, the quadratic part $S_0$ and the part with interactions $S_{int}$. 
The quadratic part is
\begin{equation}
 S_0 = \int\di^4 x\cd\ad\left[ 
  \cpsi\apsi + \cZ\frac{\dbox}{i\dpp}\aZ - iM\cZ\frac{\cd}{i\dpp}\cpsi + iM\aZ\frac{\ad}{i\dpp}\apsi 
 \right],
\end{equation}
and the interaction part is
\begin{equation}
 \begin{split}
 S_{int} &= \int\di^4 x\cd\ad\left[ 
  - i\frac{\lambda}{2}\cZ^2\frac{\cd}{i\dpp}\cpsi - i\frac{\lambda}{3}\cZ\frac{\cd}{i\dpp}\cZ\frac{i\dmp\cd}{i\dpp}\cZ 
 \right. \\ &\left.
  + i\frac{\lambda}{2}\aZ^2\frac{\ad}{i\dpp}\apsi + i\frac{\lambda}{3}\aZ\frac{\ad}{i\dpp}\aZ\frac{i\dpm\ad}{i\dpp}\aZ 
 \right].
 \end{split}
\end{equation}
Notice that the coupling constant in front of the $Z^2\psi_-$ term is the same as the coupling constant in front of the $Z^3$ term. This is a consequence of the (hidden) extra supersymmetry. When we add terms to the action which explicitly break this symmetry, there is no reason why these two coupling constants should stay the same. However, since our main aim in this paper is to construct the Feynman rules and prove that they give consistent results, for computational simplicity, in this paper we will keep the two coupling constants the same.

The extra SIM(2) invariant but not Lorentz invariant term introduced in \cite{uam} which we add to the quadratic part of the action is
\begin{equation}
S_{SIM}=im^{2}\int\di^4 x\cd\ad\left(\cZ\dfrac{1}{\dpp}\aZ\right).
\end{equation}
Then the quadratic part of the SIM(2) Wess-Zumino model becomes
\begin{equation}
 S_0 = \int\di^4 x\cd\ad\left[ 
  \cpsi\apsi + \cZ\frac{\dbox - m^{2}}{i\dpp}\aZ - iM\cZ\frac{\cd}{i\dpp}\cpsi + iM\aZ\frac{\ad}{i\dpp}\apsi 
 \right].
\end{equation}
Note that the interaction part remain the same.

\section{The generating functional}
\label{sec:genfunc}

The SIM-chiral superfields are constrained superfields, nevertheless we can define functional derivatives with respect to them, similar
to how it is done for standard chiral superfields. We define
\begin{align}
 \frac{\delta}{\delta \cl{F}(x,\theta)}\cl{F}(x',\theta') = \ad(x',\theta')\delta^4(x-x')\delta^2(\theta-\theta'), \nonumber \\
 \frac{\delta}{\delta \al{F}(x,\theta)}\al{F}(x',\theta') = \cd(x',\theta')\delta^4(x-x')\delta^2(\theta-\theta'),
\label{eq:gf:fddef}
\end{align}
where $\cl{F}$ ($\al{F}$) are (anti)chiral superfields and we have explicitly indicated on which variable the $\cd$ and $\ad$ derivatives act.
The delta function in Grassman variables is defined as $\delta^2(\theta - \theta') = \delta(\atheta^{\ai{+}}-\atheta'^{\ai{+}})\delta(\ctheta^{\ci{+}}-\ctheta'^{\ci{+}})$.
Since the right hand side of the equation is Grassman odd, the left hand side must also be Grassman odd and thus the functional derivative has to be of opposite Grassman parity as compared to the superfield with respect to which we differentiate.

Now we can try to define the generating functional. To do that we will need a source for each superfield. In our calculations
we have used an approach with chiral and antichiral supersources, so we have used two chiral supersources $\cJ$, $\ceta$ corresponding to the two chiral superfields
$\cZ$, $\cpsi$ and two antichiral supersources $\aJ$, $\aeta$ corresponding to two antichiral superfields $\aZ$, $\apsi$. 
To simplify the notation, we will denote the collection of superfields $Z$, $\bar{Z}$, $\psi_\alpha$ and $\bar{\psi}_{\dot{\alpha}}$ by $\vphi$ and the supersources $\cJ$, $\aJ$, $\ceta$, $\aeta$ by the symbol $\vJ$.

The generating functional for the free field theory will look like
\begin{equation}
 Z_0[\vJ] = \frac{1}{N_0}\int\mathcal{D}\vphi e^{ iS_0[\vphi] + i\int\vphi\vJ },
\label{eq:gf:Z0def}
\end{equation}
where $N_0 = Z[0]$ is a normalization constant and $\int\vphi\vJ$ denotes the source terms, which are of the form
\begin{equation*}
\int\vphi\vJ = \int\di^4x\cd\left(\cZ\cJ\right) - \int\di^4x\ad\left(\aZ\aJ\right) - \int\di^4x\cd\left(\cpsi\ceta\right) + \int\di^4x\ad\left(\apsi\aeta\right).
\end{equation*}
Each source term as a whole has to be Grassman even, but the integral measure is Grassman odd, so the supersource has to be of the opposite Grassman parity compared to the Grassman
parity of the superfield to which it is associated. The signs were chosen in such way that the following relations hold
\begin{align}
 \frac{\delta}{\delta \cJ(x,\theta)}\int\vphi\vJ &= \cZ(x,\theta), &
 \frac{\delta}{\delta \aJ(x,\theta)}\int\vphi\vJ &= \aZ(x,\theta), \nonumber \\
 \frac{\delta}{\delta \ceta(x,\theta)}\int\vphi\vJ &= \cpsi(x,\theta), &
 \frac{\delta}{\delta \aeta(x,\theta)}\int\vphi\vJ &= \apsi(x,\theta).
\label{eq:gf:srcder}
\end{align}
This will save us some effort when we will be replacing the superfields with the functional derivatives with respect to supersources in the interacting part of the action.
It may seem unusual to have supersources with Grassman parity opposite to the Grassman parity of the associated superfield, but there is no problem with it, because the Grassman parity of the functional derivative is opposite to the Grassman parity of the supersource with respect to which we differentiate, so the superfield and the supersource functional derivative associated with it has the same Grassman parity. The source terms can be rewritten using integrals over the full SIM-superspace as
\begin{equation*}
 \int\vphi\vJ = \int\di^4x\cd\ad\left( \cZ\frac{\cd}{i\dpp}\cJ + \aZ\frac{\ad}{i\dpp}\aJ + \cpsi\frac{\cd}{i\dpp}\ceta + \apsi\frac{\ad}{i\dpp}\aeta \right).
\end{equation*}

If we think of the collection of superfields $\vphi$ and the collection of supersources $\vJ$ as column vectors
\begin{align*}
 \vphi &= 
  \begin{pmatrix}
   \cZ \\ \aZ \\ \cpsi \\ \apsi
  \end{pmatrix}, &
 \vJ &= 
  \begin{pmatrix}
   \cJ \\ \aJ \\ \ceta \\ \aeta
  \end{pmatrix},
\end{align*}
then the expression \eqref{eq:gf:Z0def} for the generating functional can be written in a compact form as
\begin{equation*}
 Z_0[\vJ] = \frac{1}{N_0}\int\mathcal{D}\vphi e^{ i\int\di^4x\cd\ad\left( \frac{1}{2}\vphi^TQ\vphi + \vphi^T\mathfrak{D}\vJ \right) },
\end{equation*}
where the superscript $T$ denotes the usual matrix transpose and the matrices $Q$ and $\mathfrak{D}$ are
\begin{align*}
 Q &= 
 \begin{pmatrix}
  0 & \frac{\dbox-m^2}{i\dpp} & -iM\frac{\cd}{i\dpp} & 0 \\
  -\frac{\dbox-m^2}{i\dpp} & 0 & 0 & iM\frac{\ad}{i\dpp} \\
  iM\frac{\cd}{i\dpp} & 0 & 0 & 1 \\
  0 & -iM\frac{\ad}{i\dpp} & -1 & 0
 \end{pmatrix}, &
 \mathfrak{D} &= 
 \begin{pmatrix}
  \frac{\cd}{i\dpp} & 0 & 0 & 0 \\
  0 & \frac{\ad}{i\dpp} & 0 & 0 \\
  0 & 0 & \frac{\cd}{i\dpp} & 0 \\
  0 & 0 & 0 & \frac{\ad}{i\dpp}
 \end{pmatrix}.
\end{align*}
The expression in the exponential can be completed to a square by shifting the integration variables in the functional integral to the new variables $\vphi'$ related to the old ones as $\vphi = \vphi' - Q^{-1}\mathfrak{D}\vJ$. 
After completing the square, the exponential can be written as a product of a part depending on $\vphi'$ and a part which does not depend on $\vphi'$. The result of the gaussian integral 
\begin{equation*}
 \int\mathcal{D}\vphi' e^{ i\int\di^4x\cd\ad\left( \frac{1}{2}\vphi'^TQ\vphi' \right)}
\end{equation*}
can be absorbed in the normalization constant $N_0$. The remaining part can be written as
\begin{equation}\begin{split}
\label{eq:gf:Z0x}
 Z_0[\vJ] = \exp\left[ i\int\di^4x\cd\ad\left(
  - \cJ\frac{1}{\dbox-m^2-M^2}\aJ - \ceta\frac{\dbox-m^2}{i\dpp(\dbox-m^2-M^2)}\aeta
 \right.\right.\\ \left.\left.
  + iM\cJ\frac{\cd}{i\dpp(\dbox-m^2-M^2)}\ceta - iM\aJ\frac{\ad}{i\dpp(\dbox-m^2-M^2)}\aeta
 \right)\right].
\end{split}\end{equation}

The full generating functional is defined as
\begin{equation}
\label{eq:gf:Zdef}
 Z[\vJ] = \frac{1}{N_0}\int\mathcal{D}\vphi e^{ iS[\vphi] + i\int\vphi\vJ }.
\end{equation}
We separate $S[\vphi]$ into a quadradratic part $S_0[\vphi]$ and a part which is higher order in fields and thus contains the interactions $S_{int}[\vphi]$. Then we replace the superfields in $S_{int}[\vphi]$ with functional derivatives with respect to supersources by using \eqref{eq:gf:srcder} and move the part of the exponential with $S_{int}$ before the integral. The remaining path integral will be exactly the path integral in the definition of a free field generating functional \eqref{eq:gf:Z0def} so we obtain 
\begin{equation}
\label{eq:gf:Zx}
 Z[\cJ, \aJ, \ceta, \aeta] = \exp\left[iS_{int}\left( \frac{1}{i}\frac{\delta}{\delta\cJ}, \frac{1}{i}\frac{\delta}{\delta\aJ}, 
  \frac{1}{i}\frac{\delta}{\delta\ceta}, \frac{1}{i}\frac{\delta}{\delta\aeta} \right)\right] Z_0[\cJ, \aJ, \ceta, \aeta].
\end{equation}
The generating functional of connected supergraphs is defined as
\begin{equation}
\label{eq:gf:Edef}
 E[\vJ] = -i\ln Z[\vJ].
\end{equation}

For most calculations it is more convenient to work in momentum space. We need to know how the derivatives $\cd$, $\ad$ act in momentum space and which identities they fulfill.
For each function $f(x,\theta)$ we define its Fourier transformed counterpart $f(p,\theta)$, which is related to the original function in such a way, that the following relations hold
\begin{align*}
 f(p,\theta) &= \int\di^4 xf(x,\theta)e^{ipx}, &
 f(x,\theta) &= \int\frac{\di^4 p}{(2\pi)^4}f(p,\theta)e^{-ipx},
\end{align*}
we will use the same symbol for both the original function and its Fourier transform, but it should be clear from the arguments which function we have in mind. We also need to know how the the derivatives acting on the original function affect its Fourier transform
\begin{align*}
 i\pdd{\alpha}{\alpha} f(x) &\quad\rightarrow\quad \mdd{p}{\alpha}{\alpha} f(p), \\
 \dbox f(x) &\quad\rightarrow\quad -p^2 f(p), \\
 \cd(x,\theta) f(x,\theta) &\quad\rightarrow\quad \cd(p, \theta) f(p,\theta) = (\cl{\partial}_{\ci{+}} + \frac{1}{2}\athetau{+}\mpp{p}) f(p,\theta), \\
 \ad(x,\theta) f(x,\theta) &\quad\rightarrow\quad \ad(p, \theta) f(p,\theta) = (\al{\partial}_{\ai{+}} + \frac{1}{2}\cthetau{+}\mpp{p}) f(p,\theta) 
\end{align*}
The derivatives $\cd$ $\ad$ fulfills anticommutation relation 
\begin{equation*}
 \{\cd(p, \theta),\ad(p,\theta)\} = \mpp{p},
\end{equation*}
which leads to the identities 
\begin{align}
\label{eq:ft:dalgebra}
 \cd(p,\theta)\ad(p,\theta)\cd(p,\theta) &= \mpp{p}\cd(p,\theta),  \\
 \ad(p,\theta)\cd(p,\theta)\ad(p,\theta) &= \mpp{p}\ad(p,\theta). \nonumber
\end{align}

To see that the momentum space spinor derivatives satify the graded Leibniz rule one has to take into account that the momentum dependence of the derivative is correlated with the momentum dependence of the field it acts on. I.e., using the symbol $\mathcal{D}$ for
any of the spinor derivatives, we have
\begin{equation}
\label{eq:ft:leibniz}
 \mathcal{D}(p+q,\theta)(f\cdot g) = \mathcal{D}(p,\theta)f\cdot g + (-1)^{\gp{f}}f\cdot\mathcal{D}(q,\theta)g.
\end{equation}
This makes it possible to integrate by parts
\begin{equation}
\label{eq:ft:perpartes}
 \int\cd\ad\left[ f\cdot\mathcal{D}(p,\theta)g \right] = - (-1)^{\gp{f}}\int\cd\ad\left[ \mathcal{D}(-p,\theta)f\cdot g \right].
\end{equation}
Another useful identity which can be proven by direct calculation is
\begin{equation}
\label{eq:ft:ddelta}
 \mathcal{D}(p,\theta)\delta^2(\theta - \theta') = -\mathcal{D}(-p,\theta')\delta^2(\theta - \theta').
\end{equation}

Finally we may write the Fourier transformed generating functional. The quadratic part of the generating functional becomes
\begin{equation}\begin{split}
\label{eq:gf:Z0p}
 Z_0[\vJ] = \exp\biggl[ i\int\frac{\di^4p}{(2\pi)^4}\cd\ad\biggl(
  &- \cJ(-p,\theta)\frac{1}{-p^2-m^2-M^2}\aJ(p,\theta) \\
  &- \ceta(-p,\theta)\frac{-p^2-m^2}{\mpp{p}(-p^2-m^2-M^2)}\aeta(p,\theta) \\
  &+ iM\cJ(-p,\theta)\frac{\cd(p,\theta)}{\mpp{p}(-p^2-m^2-M^2)}\ceta(p,\theta) \\
  &- iM\aJ(-p,\theta)\frac{\ad(p,\theta)}{\mpp{p}(-p^2-m^2-M^2)}\aeta(p,\theta)
 \biggr)\biggr].
\end{split}\end{equation}
The expression for the full generating functional will look like \eqref{eq:gf:Zx} with $S_{int}$ given by
\begin{equation*}\begin{split}
S_{int}[\vphi] = \int\frac{\di^4p_1}{(2\pi)^4}&\frac{\di^4p_2}{(2\pi)^4}\frac{\di^4p_3}{(2\pi)^4}\cd\ad\biggl[(2\pi)^4\delta^4(p_1+p_2+p_3)\biggl( \\
  &- i\frac{\lambda}{2}\cZ(p_1,\theta)\cZ(p_2,\theta)\frac{\cd(p_3,\theta)}{\mpp{p_3}}\cpsi(p_3,\theta) \\
  &- i\frac{\lambda}{3}\cZ(p_1,\theta)\frac{1}{\mpp{p_2}}\cd(p_2,\theta)\cZ(p_2,\theta)\frac{\mmp{p_3}}{\mpp{p_3}}\cd(p_3,\theta)\cZ(p_3,\theta) \\
  &+ i\frac{\lambda}{2}\aZ(p_1,\theta)\aZ(p_2,\theta)\frac{\ad(p_3,\theta)}{\mpp{p_3}}\apsi(p_3,\theta) \\
  &+ i\frac{\lambda}{3}\aZ(p_1,\theta)\frac{1}{\mpp{p_2}}\ad(p_2,\theta)\aZ(p_2,\theta)\frac{\mpm{p_3}}{\mpp{p_3}}\ad(p_3,\theta)\aZ(p_3,\theta)
 \biggr)\biggr].
\end{split}\end{equation*}
where we have to remember that $\phi(p)$ is replaced by
\begin{equation*}
 \vphi(p,\theta) \quad\rightarrow\quad \frac{(2\pi)^4}{i}\frac{\delta}{\delta\vJ(-p,\theta)},
\end{equation*}
which can be deduced from the Fourier transformed source term $\int\vphi\vJ$. 
 
\section{Feynman rules}
\label{sec:fr}

\newcommand{\gi}[1]{^{[#1]}}

In this section the Feynman rules of our SIM supersymmetric field theory will be derived. Rather than just state the rules we will derive them by calculating a simple one-loop two-point supergraph using the path integral.
This will give us insight into the structure of the expressions with which we will have to deal. Then we will use this result to calculate the contribution to the effective action, i.e. amputate external propagators and replace them with superfields. During the calculation it will be shown how certain expressions can be depicted in the form of supergraphs and at the end we will be able to state the Feynman rules and describe how the supergraph expressions can be assembled from propagators, vertices and external superfields. This gives us the ability to easily create a mathematical expression for any supergraph contributing to the effective action. The manipulations which can be performed directly on the supergraphs (d-algebra) will be described in the next section.

\subsection{From the path integral to the amplitude}
Our first task will be finding an expression for a supergraph 
\parbox{15mm}{
\begin{fmfgraph*}(15,6)
 \fmfleft{i} \fmfright{o} \fmf{plain}{i,c} \fmf{plain}{a,o} \fmf{plain,left,tension=.5}{c,a,c}
\end{fmfgraph*}}
contributing to $\frac{\delta^2 E[\vJ]}{\delta\cJ\delta\aJ}|_{\vJ=0}$, with vertex on the left of the type $\cZ\cZ\cpsi$, the vertex on the right of the type $\aZ\aZ\apsi$, both external propagators and one propagator in the loop of type $\cZ-\aZ$, and the other propagator in the loop of type $\cpsi-\apsi$. To create such an expression we will expand the exponentials in \eqref{eq:gf:Zx} and \eqref{eq:gf:Z0p}. We will be interested only in the the terms important for calculating our supergraph, to simplify things we will also not include the factors $\frac{1}{n!}$ coming from the Taylor expansion of exponentials. This gives us the expression
\begin{equation}\begin{split}
\label{eq:fr:Eone}
 -i \frac{\delta}{\delta\cJ(-p',\theta')} \frac{\delta}{\delta\aJ(-p,\theta)}
 &\left[ i\frac{\lambda}{2}\int\di^4p_1\di^4p_2\di^4p_3\di^2\theta_A(2\pi)^4\delta^4(p_1+p_2+p_3) \right. \\ &\qquad\left.
  \times \frac{\delta}{\delta\cJ(-p_1,\theta_A)}\frac{\delta}{\delta\cJ(-p_2,\theta_A)}
  \frac{\cd\gi{1}(p_3,\theta_A)}{\mpp{p_3}}\frac{\delta\gi{2}}{\delta\ceta(-p_3,\theta_A)} \right] \\
 &\left[ -i\frac{\lambda}{2}\int\di^4q_1\di^4q_2\di^4q_3\di^2\theta_B(2\pi)^4\delta^4(q_1+q_2+q_3) \right. \\ &\qquad\left.
  \times \frac{\delta}{\delta\aJ(-q_1,\theta_B)}\frac{\delta}{\delta\aJ(-q_2,\theta_B)}
  \frac{\ad\gi{3}(q_3,\theta_B)}{\mpp{q_3}}\frac{\delta\gi{4}}{\delta\aeta(-q_3,\theta_B)} \right] \\
 &\left[ -i\int\frac{\di^4k_1}{(2\pi)^4}\di^2\theta_1\ceta(-k_1,\theta_1)\frac{-k_1^2-m^2}{\mpp{k_1}(-k_1^2-m^2-M^2)}\aeta(k_1,\theta_1) \right] \\
 &\left[ -i\int\frac{\di^4k_2}{(2\pi)^4}\di^2\theta_2\cJ\gi{5}(-k_2,\theta_2)\frac{1}{-k_2^2-m^2-M^2}\aJ\gi{6}(k_2,\theta_2) \right] \\
 &\left[ -i\int\frac{\di^4k_3}{(2\pi)^4}\di^2\theta_3\cJ\gi{7}(-k_3,\theta_3)\frac{1}{-k_3^2-m^2-M^2}\aJ\gi{8}(k_3,\theta_3) \right] \\
 &\left[ -i\int\frac{\di^4k_4}{(2\pi)^4}\di^2\theta_4\cJ\gi{9}(-k_4,\theta_4)\frac{1}{-k_4^2-m^2-M^2}\aJ\gi{10}(k_4,\theta_4) \right],
\end{split}\end{equation}
where we have used the shorthand notation $\di^2\theta$ for the integral measure $\cd\ad$.
The $-i$ factor at the beginning comes from the fact that we have $E[\vJ]=-i\ln Z[\vJ]$. The two initial functional derivatives give us the propagators of the external legs, then the terms in the first two brackets come from the two vertices and the four last brackets each give a propagator.
Notice that all Grassman odd objects have been given numbers. This will help us to keep track of minus signs coming from their anticommutation properties. 
The eight functional derivatives each act on one of the eight supersources, so so that there are no supersources left in the final expression. There are $36$ different ways that the functional derivatives may act, the final result will be a combination of all these terms. Each of the individual contributions can be represented as a supergraph. We will consider only one particle irreducible supergraphs for the calculation of the effective action. To illustrate our methods, we choose one of the supergraphs in the set of terms we get from the expression given above. There might be several terms corresponding to the same supergraph but that fact, together with the $1/n!$ in front of the term, will be taken into account later when we compute the symmetry factor
\footnote{
In our case there is a $\frac{1}{3!}$ factor from the expansion of the exponential in $Z_0$ and the number of terms corresponding to the same supergraph is $24$, so the symmetry factor would be $\frac{24}{3!}$.}.
Let us choose the term which can be written as
\begin{equation}\begin{split}
\label{eq:fr:Etwo}
 &-i\left(i\frac{\lambda}{2}\right)\left(-i\frac{\lambda}{2}\right)
  \int\di^4p_1\di^4p_2\di^4p_3\di^2\theta_A\int\di^4q_1\di^4q_2\di^4q_3\di^2\theta_B \\
 &(2\pi)^4\delta^4(p_1+p_2+p_3)(2\pi)^4\delta^4(q_1+q_2+q_3) \\
 &\left[ \frac{\delta}{\delta\cJ(-p_2,\theta_A)}\frac{\delta}{\delta\aJ(-q_2,\theta_B)}
  \int\frac{\di^4k_2}{(2\pi)^4}\di^2\theta_2\cJ\gi{5}(-k_2,\theta_2)\frac{-i}{-k_2^2-m^2-M^2}\aJ\gi{6}(k_2,\theta_2) \right] \\
 &\left[ \frac{\cd\gi{1}(p_3,\theta_A)}{\mpp{p_3}}\frac{\delta\gi{2}}{\delta\ceta(-p_3,\theta_A)}
  \frac{\ad\gi{3}(q_3,\theta_B)}{\mpp{q_3}}\frac{\delta\gi{4}}{\delta\aeta(-q_3,\theta_B)} \right. \\ &\qquad\qquad\qquad\qquad\left.
  \int\frac{\di^4k_1}{(2\pi)^4}\di^2\theta_1\ceta(-k_1,\theta_1)\frac{-i(-k_1^2-m^2)}{\mpp{k_1}(-k_1^2-m^2-M^2)}\aeta(k_1,\theta_1) \right] \\
 &\left[ \frac{\delta}{\delta\cJ(-p',\theta')}\frac{\delta}{\delta\aJ(-q_1,\theta_B)} 
  \int\frac{\di^4k_3}{(2\pi)^4}\di^2\theta_3\cJ\gi{7}(-k_3,\theta_3)\frac{-i}{-k_3^2-m^2-M^2}\aJ\gi{8}(k_3,\theta_3) \right] \\
 &\left[ \frac{\delta}{\delta\cJ(-p_1,\theta_A)}\frac{\delta}{\delta\aJ(-p,\theta)} 
  \int\frac{\di^4k_4}{(2\pi)^4}\di^2\theta_3\cJ\gi{9}(-k_4,\theta_4)\frac{-i}{-k_4^2-m^2-M^2}\aJ\gi{10}(k_4,\theta_4) \right].
\end{split}\end{equation}
To get this expression we had to anticommute several Grassaman odd variables. Since in SIM-superspace there are no individual spinor indices on Grassman objects we needed to introduce individual
numbers that will label each Grassman odd object. Then we may compute the sign we get from anticommutation by looking at the permutation of these numbers. If the permutation is odd, there is an extra minus sign.
In our case the permutation is $\left(\begin{smallmatrix}1&2&3&4&5&6&7&8&9&10\\5&6&1&2&3&4&7&8&9&10\end{smallmatrix}\right)$ which is even so we have no extra minus sign.
The expression contains four subexpressions, which all take the following general form
\begin{equation}
\label{eq:fr:VPgen}
 \mathcal{O}_A(p_A,\theta_A)\frac{\delta}{\delta\vJ_A(-p_A,\theta_A)} \cdot \mathcal{O}_B(p_B,\theta_B)\frac{\delta}{\delta\vJ_B(-p_B,\theta_B)}
   \cdot \int\di^4p\di^2\theta\vJ_A(-p,\theta)\mathcal{P}(p,\theta)\vJ_B(p,\theta)
\end{equation}
where $\mathcal{O}_A(p,\theta)$, $\mathcal{O}_B(p,\theta)$ and $\mathcal{P}(p,\theta)$ are expressions containing $p$, $\theta$ and derivatives with respect to $\theta$. Moreover, a Grassman parity is assigned to each of these expressions. The $\vJ_A$ and $\vJ_B$ is a collective notation for the supersources and the corresponding superfields are collectively denoted by $\vphi_A$ and $\vphi_B$.

When we do the functional derivatives, we have to keep in mind that some of the objects are Grassman odd and some of them Grassman even, so there may be a sign factor depending on the grassman parity of $\vJ_A$, $\vJ_B$, $\mathcal{O}_A$, $\mathcal{O}_B$, $\mathcal{P}$. But for each of the possible combination of Grassman parities of $\vJ_A$, $\vJ_B$ the result can be written in a form, that does not depend on the grassman parities of $\mathcal{O}_A$, $\mathcal{O}_B$ and $\mathcal{P}$.
The resulting expression consists of operators $\mathcal{O}_A$, $\mathcal{O}_B$, $\mathcal{P}$ and covariant derivatives from chiral functional derivatives acting on $\delta^2(\theta_A-\theta_B)\delta^4(p_A+p_B)$. The sign will be
automatically taken care of if we write the operators in the same order as in
the original expression \eqref{eq:fr:VPgen} but with the covariant derivatives coming from the chiral functional derivatives in the position of the functional derivative if the functional derivative is Grassman odd but in the position of the source if the functional derivative is Grassman  even.
For example if $\vphi_A$ is Grassman even and $\vphi_B$ is Grassman odd then the result is
\begin{equation}
\label{eq:fr:VPeo}
 \mathcal{O}_A(p_A,\theta_A)\cdot\mathcal{O}_B(p_B,\theta_B)\mathcal{D}_B(p_B,\theta_B)\cdot\mathcal{D}_A(p_A,\theta_A)\mathcal{P}(p_A,\theta_A)
 \delta^2(\theta_A-\theta_B)\delta^4(p_A+p_B),
\end{equation}
where $\mathcal{D}_A$($\mathcal{D}_B$) is $\ad$ if the supersource is chiral and $\cd$ if the supersource is antichiral. Note also that $\mathcal{P}$ is in the variables $p_A,\theta_A$.


Using the above rules the \eqref{eq:fr:Etwo} can be written as
\begin{equation}\begin{split}
 &-i\left(i\frac{\lambda}{2}\right)\left(-i\frac{\lambda}{2}\right)\frac{1}{[(2\pi)^4]^2}\int\frac{\di^4q}{(2\pi)^4}\di^2\theta_A\di^2\theta_B
  (2\pi)^4\delta^4(p+p') \\
 &\ad\gi{5}(p-q,\theta_A)\frac{-i}{-(p-q)^2-m^2-M^2}\cd\gi{6}(q-p,\theta_B)\delta^2(\theta_A-\theta_B) \\
 &\frac{\cd\gi{1}(q,\theta_A)}{\mpp{q}}\ad\gi{2}(q,\theta_A)\frac{\ad\gi{3}(-q,\theta_B)}{-\mpp{q}}\cd\gi{4}(-q,\theta_B)
  \frac{(-i)(-q^2-m^2)}{\mpp{q}(-q^2-m^2-M^2)}\delta^2(\theta_A-\theta_B) \\
 &\ad\gi{7}(-p,\theta)\frac{-i}{-p^2-m^2-M^2}\cd\gi{8}(p,\theta_B)\delta^2(\theta-\theta_B) \\
 &\ad\gi{9}(-p,\theta_A)\frac{-i}{-p^2-m^2-M^2}\cd\gi{10}(p,\theta')\delta^2(\theta'-\theta_A).
\end{split}\end{equation}

\subsection{From the amplitude to the supergraph}
The result from the previous subsection can be represented as a Feynman supergraph. We will first give the result and then we state the general rules and explain how the supergraph correspond to the mathematical expression given above.
\newcommand{\fmfopext}[5]{
 \fmfi{phantom,label=\small #4,label.dist=1,label.side=#3,label.dist=#5}{subpath ((#1-0.01)*length(#2),(#1+0.01)*length(#2)) of #2}
}
\newcommand{\fmfop}[4]{
 \fmfopext{#1}{#2}{#3}{#4}{1}
}
\newcommand{\fmfarrowpl}[4]{
 \fmfi{plain,width=2,label=\small #4,label.dist=10,label.side=left,#3}{subpath ((#1-#2)*length(#3),(#1+#2)*length(#3)) of #3 shifted 45 dir(angle(direction #1*length(#3) of #3)+90)}
 \fmfcmd{%
  path p;
  p:=subpath ((#1-#2)*length(#3),(#1+#2)*length(#3)) of #3 shifted 45 dir(angle(direction #1*length(#3) of #3)+90);
  shrink(0.7);
  cfill(harrow(p,length(p)));
  endshrink;
 }
}
\newcommand{\fmfarrownl}[4]{
 \fmfi{plain,width=2,label=\small #4,label.dist=10,label.side=left,#3}{subpath ((#1-#2)*length(#3),(#1+#2)*length(#3)) of #3 shifted 45 dir(angle(direction #1*length(#3) of #3)+90)}
 \fmfcmd{%
  path p;
  p:=subpath ((#1-#2)*length(#3),(#1+#2)*length(#3)) of #3 shifted 45 dir(angle(direction #1*length(#3) of #3)+90);
  shrink(0.7);
  cfill(harrow(reverse p,length(p)));
  endshrink;
 }
}
\newcommand{\fmfarrowpr}[4]{
 \fmfi{plain,width=2,label=\small #4,label.dist=10,label.side=right,#3}{subpath ((#1-#2)*length(#3),(#1+#2)*length(#3)) of #3 shifted 45 dir(angle(direction #1*length(#3) of #3)+270)}
 \fmfcmd{%
  path p;
  p:=subpath ((#1-#2)*length(#3),(#1+#2)*length(#3)) of #3 shifted 45 dir(angle(direction #1*length(#3) of #3)+270);
  shrink(0.7);
  cfill(harrow(p,length(p)));
  endshrink;
 }
}
\newcommand{\fmfarrownr}[4]{
 \fmfi{plain,width=2,label=\small #4,label.dist=10,label.side=right,#3}{subpath ((#1-#2)*length(#3),(#1+#2)*length(#3)) of #3 shifted 45 dir(angle(direction #1*length(#3) of #3)+270)}
 \fmfcmd{%
  path p;
  p:=subpath ((#1-#2)*length(#3),(#1+#2)*length(#3)) of #3 shifted 45 dir(angle(direction #1*length(#3) of #3)+270);
  shrink(0.7);
  cfill(harrow(reverse p,length(p)));
  endshrink;
 }
}
\newcommand{\fmffix}[4]{
 \fmfforce{point #1*length(#3) of #3 shifted #2 dir(angle(direction #1*length(#3) of #3)+270)}{#4}
}
\newcommand{\fmftext}[3]{
 \fmfi{phantom,#3}{subpath ((#1-0.01)*length(#2),(#1+0.01)*length(#2)) of #2}
}
\begin{equation}\begin{split}
\label{eq:fg:extfirst}
&-i\left(i\frac{\lambda}{2}\right)\left(-i\frac{\lambda}{2}\right)(-i)^4\times \\
&\quad\begin{fmfgraph*}(140,53)
 \fmfv{label=$\theta'$}{i} \fmfleft{i}
 \fmfv{label=$\theta$}{o} \fmfright{o}
 \fmfv{label=$\theta_A$,label.angle=0}{c}
 \fmfv{label=$\theta_B$,label.angle=180}{a}
 \fmf{plain}{c,i} \fmf{plain}{o,a} \fmf{plain,left,tension=.5,tag=1}{c,a} \fmf{plain,left,tension=.5,tag=2}{a,c}
 \fmffreeze
 \fmfop{0.15}{vpath(__i,__c)}{right}{$\ad\gi{9}$}
 \fmfop{0.65}{vpath(__i,__c)}{right}{$\cfrac{1}{\dbox-m^2-M^2}$}
 \fmfop{0.95}{vpath(__i,__c)}{right}{$\cd\gi{10}$}
 \fmfop{0.05}{vpath(__o,__a)}{right}{$\ad\gi{7}$}
 \fmfop{0.38}{vpath(__o,__a)}{right}{$\cfrac{1}{\dbox-m^2-M^2}$}
 \fmfop{0.85}{vpath(__o,__a)}{right}{$\cd\gi{8}$}
 \fmfop{0.15}{vpath1(__c,__a)}{left}{$\ad\gi{5}$}
 \fmfop{0.3}{vpath1(__c,__a)}{left}{$\cfrac{1}{\dbox-m^2-M^2}$}
 \fmfop{0.85}{vpath1(__c,__a)}{left}{$\cd\gi{6}$}
 \fmfop{0.12}{vpath2(__a,__c)}{left}{$\cfrac{\ad\gi{3}}{i\dpp}$}
 \fmfop{0.22}{vpath2(__a,__c)}{left}{$\cd\gi{4}$}
 \fmfop{0.64}{vpath2(__a,__c)}{left}{$\cfrac{\dbox-m^2}{i\dpp(\dbox-m^2-M^2)}$}
 \fmfop{0.78}{vpath2(__a,__c)}{left}{$\ad\gi{2}$}
 \fmfop{0.88}{vpath2(__a,__c)}{left}{$\cfrac{\cd\gi{1}}{i\dpp}$}
 \fmfarrowpl{0.5}{0.1}{vpath(__c,__i)}{$p$}
 \fmfarrowpl{0.5}{0.1}{vpath(__a,__o)}{$p$}
 \fmfarrownr{0.5}{0.1}{vpath1(__c,__a)}{$p-q$}
 \fmfarrowpr{0.5}{0.1}{vpath2(__c,__a)}{$q$}
\end{fmfgraph*} \\
\end{split}\end{equation}
The supergraph consists of lines (propagators) either connecting vertices or coming out of the graph as an external leg. A $\theta$ variable is assigned to each vertex and to the open end of each external leg so all propagators come equipped with $\theta$ variables at both ends. Moreover a momentum flow is assigned to each propagator in such a way that it is conserved at each vertex. An empty propagator (without any attached objects) comes with a delta function in the $\theta$ variables at its ends, so we may write
\begin{equation}
 \parbox{50mm}{
  \begin{fmfgraph*}(40,10)
   \fmfv{label=$\theta$}{i} \fmfleft{i} \fmfv{label=$\theta'$}{o} \fmfright{o} \fmf{plain}{i,o}  
   \fmfarrownr{0.5}{0.1}{vpath(__i,__o)}{p}
  \end{fmfgraph*}
 } = \delta^2(\theta-\theta').
\end{equation}
If there are some operators acting on the delta function, then we write them at the propagator. When doing so we have to take care of a few things. We have to distinguish to which end the operators belong. We use the convention that all operators act on the delta function from the left. If there are multiple operators then we have to write them in such a way that if some operator is placed closer to the end of the propagator relative to some other operator, then it is written to the left of the other operator (further from the delta function). 
We interchangeably use configuration and momentum space expressions where we understand them to be related as
\begin{align}
 i\pdd{\alpha}{\alpha}&\rightarrow\mdd{p}{\alpha}{\alpha}, & \dbox&\rightarrow -p^2, & \cd&\rightarrow\cd(p,\theta), & \ad&\rightarrow\ad(p,\theta),
\end{align}
where $\theta$ is variable corresponding to the end of the propagator where the operator is placed. If the momentum arrow points to the end where the operator is placed then $p$ is equal to the momentum assigned to this arrow, if the momentum arrow points in the opposite direction then $p$ is equal to minus the momentum.
For example the bottom half circle in the loop in the supergraph \eqref{eq:fg:extfirst} corresponds to
\begin{equation*}
 \frac{\cd\gi{1}(q,\theta_A)}{\mpp{q}}\ad\gi{2}(q,\theta_A)\frac{-q^2-m^2}{\mpp{q}(-q^2-m^2-M^2)}
 \frac{\ad\gi{3}(-q,\theta_B)}{-\mpp{q}}\cd\gi{4}(-q,\theta_B)
 \delta^2(\theta_A-\theta_B).
\end{equation*}
Note that it is not important if the operators in the $\theta_A$ variables are written before the operators in the $\theta_B$ variables or vice versa.
To finish the expression corresponding to the Feynman supergraph we have to add $\int\frac{\di^4q}{(2\pi)^4}$ for each loop, $\int\di^2\theta$ for each vertex,  $\frac{1}{(2\pi)^4}$ for each external leg and the overall momentum conservation delta function multiplied by $(2\pi)^4$. The last thing which we have to do is to look at the expression we have written down and determine if the sequence of numbers labeling Grassman odd objects is an even or an odd permutation. If it is odd we have to multiply the whole expression with a factor $-1$.

\subsection{From connected diagrams to the effective action}
The rules as stated above will give us the connected correlation functions. However, usually we would like to calculate contributions to the effective action. This means that we have to consider only one particle irreducible supergraphs and we have to amputate the external leg propagators and attach superfields.

The propagators all come from expressions of the type \eqref{eq:fr:VPgen} which in the case when it is an external propagator looks like
\begin{equation}
\label{eq:fr:VEgen}
 \frac{\delta}{\delta\vJ_{ext}(-p_{ext},\theta_{ext})} \cdot \mathcal{O}_A(p_A,\theta_A)\frac{\delta}{\delta\vJ_A(-p_A,\theta_A)}
  \cdot \int\di^4p\di^2\theta\vJ_{ext}(-p,\theta)\mathcal{P}(p,\theta)\vJ_A(p,\theta).
\end{equation}
since there is never an operator ${\cal O}$ acting at the external end of the external propagator. This means that removing an external propagator, we have to keep the $\cal O_A$ operator but remove everything else including the covariant derivatives coming from the fact that the sources are SIM-chiral superfields giving us the new expression
\begin{equation}
\label{eq:fr:VEres}
 c_A\mathcal{O}_A(p_A,\theta_A)\vphi_A(p_A,\theta_A),
\end{equation}
where $c_A=c_{\cZ},c_{\aZ},c_{\cpsi},c_{\apsi}$ are yet undetermined constants. It can be shown that in our setup, all of them are equal to $i$. 

In the definition of the effective action we include an explicit factor
\begin{equation*}
 \frac{1}{(\#\cZ)!(\#\aZ)!(\#\cpsi)!(\#\apsi)!},
\end{equation*}
where $\#\cZ$, $\#\aZ$, $\#\cpsi$, $\#\apsi$ is the number of external legs ending with superfield $\cJ$, $\aJ$, $\cpsi$, $\apsi$.

The contribution to the effective action corresponding to our supergraph is then equal to
\begin{equation}\begin{split}
 &-i\left(i\frac{\lambda}{2}\right)\left(-i\frac{\lambda}{2}\right)i^2\int\frac{\di^4p}{(2\pi)^4}\frac{\di^4q}{(2\pi)^4}\di^2\theta_A\di^2\theta_B
  \cZ(-p,\theta_A)\aZ(p,\theta_B) \\
 &\ad\gi{5}(p-q,\theta_A)\frac{-i}{-(p-q)^2-m^2-M^2}\cd\gi{6}(q-p,\theta_B)\delta^2(\theta_A-\theta_B) \\
 &\frac{\cd\gi{1}(q,\theta_A)}{\mpp{q}}\ad\gi{2}(q,\theta_A)\frac{\ad\gi{3}(-q,\theta_B)}{-\mpp{q}}\cd\gi{4}(-q,\theta_B)
  \frac{(-i)(-q^2-m^2)}{\mpp{q}(-q^2-m^2-M^2)}\delta^2(\theta_A-\theta_B).
\end{split}\end{equation}
Thus we see that the expression corresponding to an external leg will just be the superfield with momentum parameter equal to the momentum assigned to the momentum flow arrow if it points away from the superfield  and to minus this momentum if the arrow points to the superfield, and the $\theta$ parameter equal to the $\theta$ variable at the vertex to which it is attached. So we may write
\begin{equation*}
 \parbox{40mm}{
  \begin{fmfgraph*}(30,10)
   \fmfv{label=$\vphi_A$}{i} \fmfleft{i} \fmfv{label=$\theta$}{o} \fmfright{o} \fmf{plain}{i,o}  
   \fmfarrowpr{0.5}{0.2}{vpath(__i,__o)}{p}
  \end{fmfgraph*}
 } = \vphi_A(p,\theta).
\end{equation*}
If there are some operators placed at the external leg, then they will act on this superfield. The rules for writing them down will be the same as in the case of propagators with the only differences that they act on the superfield instead of on the delta function.

In the effective action, there should be momentum integrals over all momentum variables appearing in the expression, to achieve this we have to add $n-1$ momentum integrals $\int\frac{\di^4p}{(2\pi)^4}$ for $n$ external legs in addition to the momentum integrals coming from loops.

The Feynman supergraph for our contribution to the effective action is
\begin{equation}\begin{split}
\label{eq:fg:efffirst}
&-i\left(i\frac{\lambda}{2}\right)\left(-i\frac{\lambda}{2}\right)(-i)^2i^2\times \\
&\hspace{20mm}\parbox{100mm}{\begin{fmfgraph*}(100,52)
 \fmfv{label=$\cZ$}{i} \fmfleft{i}
 \fmfv{label=$\aZ$}{o} \fmfright{o}
 \fmfv{label=$\theta_A$,label.angle=0}{c}
 \fmfv{label=$\theta_B$,label.angle=180}{a}
 \fmf{plain}{c,i} \fmf{plain}{o,a} \fmf{plain,left,tension=.1,tag=1}{c,a} \fmf{plain,left,tension=.5,tag=2}{a,c}
 \fmffreeze
 \fmfop{0.15}{vpath1(__c,__a)}{left}{$\ad\gi{5}$}
 \fmfop{0.3}{vpath1(__c,__a)}{left}{$\cfrac{1}{\dbox-m^2-M^2}$}
 \fmfop{0.85}{vpath1(__c,__a)}{left}{$\cd\gi{6}$}
 \fmfop{0.12}{vpath2(__a,__c)}{left}{$\cfrac{\ad\gi{3}}{i\dpp}$}
 \fmfop{0.22}{vpath2(__a,__c)}{left}{$\cd\gi{4}$}
 \fmfop{0.64}{vpath2(__a,__c)}{left}{$\cfrac{\dbox-m^2}{i\dpp(\dbox-m^2-M^2)}$}
 \fmfop{0.78}{vpath2(__a,__c)}{left}{$\ad\gi{2}$}
 \fmfop{0.88}{vpath2(__a,__c)}{left}{$\cfrac{\cd\gi{1}}{i\dpp}$}
 \fmfarrowpl{0.5}{0.2}{vpath(__c,__i)}{$p$}
 \fmfarrowpl{0.5}{0.2}{vpath(__a,__o)}{$p$}
 \fmfarrownr{0.5}{0.1}{vpath1(__c,__a)}{$p-q$}
 \fmfarrowpr{0.5}{0.1}{vpath2(__c,__a)}{$q$}
\end{fmfgraph*}}\quad
\end{split}\end{equation}

\subsection{From the supergraph to the amplitude}
So far, we have calculated a contribution to the effective action directly from the path integral and we have shown how to associate a supergraph to it. In the following we will construct rules that will make it possible to start with the supergraph and find the mathematical expression for the contribution to the effective action from it. The calculations are similar to supergraph calculations in ordinary superspace with the main difference being the need to number the individual Grassman objects to keep track of signs.

We start by describing the method for numbering Grassman objects in supergraphs. These objects can be either external fields or operators. 
The operators come from three sources; they can come from the quadratic part or the interaction part of the action as well as from the functional differentiation of chiral sources. We divide the Grassman objects into two classes. In the first class we put everything coming from the interaction part of the action and it will be associated with the vertices. In the second class we put everything coming from the quadratic part of the action and it will be associated with the propagators. External fields will be associated with the vertices. Finally, the Grassman odd operators coming from the functional differentiation of chiral sources will be associated with vertices if the source is Grassman even and with the propagator otherwise.

We will now number the Grassman objects in each vertex according to the following rule. A general vertex comes from a term in the interaction part of the action of the following form
\begin{equation*}
 K\int\di^4x\cd\ad\left(\mathcal{O}_{\cZ}\cZ\cdot\mathcal{O}_{\aZ}\aZ\cdot\mathcal{O}_{\cpsi}\cpsi\cdot\mathcal{O}_{\apsi}\apsi\cdots\right),
\end{equation*}
where $K$ is some constant and $\mathcal{O}_{\cZ}$, $\mathcal{O}_{\aZ}$, $\mathcal{O}_{\cpsi}$, $\mathcal{O}_{\apsi}$ are some operators. We can number the Grassman odd objects appearing in this expression from left to right (the same thing as we did in \eqref{eq:fr:Eone}). This can be represented graphically as
\vspace{2mm}
\begin{equation}
i^{1-n}K\times\qquad\parbox{40mm}{\begin{fmfgraph*}(40,40)
 \fmfv{label=$\cZ$}{cz} \fmfv{label=$\aZ$}{az} \fmfv{label=$\cpsi$}{cp} \fmfv{label=$\apsi$}{ap}
 \fmfsurround{ap,cp,az,cz,d0}
 \fmfv{}{v} \fmfdot{v} \fmf{plain}{cz,v} \fmf{plain}{az,v} \fmf{plain}{cp,v} \fmf{plain}{ap,v} \fmf{phantom}{d0,v}
 \fmffreeze
 \fmfv{}{d1,d2,d3} \fmffix{0.3}{0}{vpath(__v,__cz)}{d1} \fmffix{0.3}{0}{vpath(__v,__ap)}{d2} \fmffix{0.3}{0}{vpath(__v,__d0)}{d3} 
 \fmfipath{ang}
 \fmfiset{ang}{vloc(__d1)..vloc(__d3)..vloc(__d2)}
 \begin{fmffor}{xang}{0.3}{0.2}{0.9}
  \fmfiv{decor.shape=circle,decor.filled=full,decor.size=thick}{point xang*length(ang) of ang}
 \end{fmffor}
 \fmfopext{0.5}{vpath(__v,__cz)}{right}{$\mathcal{O}_{\cZ}$}{10}
 \fmfopext{0.5}{vpath(__v,__az)}{right}{$\mathcal{O}_{\aZ}$}{10}
 \fmfop{0.5}{vpath(__v,__cp)}{right}{$\mathcal{O}_{\cpsi}$}
 \fmfopext{0.2}{vpath(__v,__cp)}{right}{$\al{\mathcal{D}}$}{25}
 \fmfop{0.5}{vpath(__v,__ap)}{right}{$\mathcal{O}_{\apsi}$}
 \fmfopext{0.2}{vpath(__v,__ap)}{right}{$\cl{\mathcal{D}}$}{25}
 \fmfv{}{fb,fa,f0,f1,f2,f3,f4,f5,f6,f7,f8,f9,f10,f11,f12,f13}
 \fmffix{0.8}{140}{vpath(__v,__cz)}{fb}
 \fmffix{0.1}{140}{vpath(__v,__cz)}{fa}
 \fmffix{0.9}{30}{reverse vpath(__v,__cz)}{f0}
 \fmffix{0.5}{30}{reverse vpath(__v,__cz)}{f1}
 \fmffix{0.5}{140}{vpath(__v,__az)}{f2}
 \fmffix{0.2}{140}{vpath(__v,__az)}{f3}
 \fmffix{0.9}{30}{reverse vpath(__v,__az)}{f4}
 \fmffix{0.7}{30}{reverse vpath(__v,__az)}{f5}
 \fmffix{0.6}{150}{vpath(__v,__cp)}{f6}
 \fmffix{0.2}{150}{vpath(__v,__cp)}{f7}
 \fmffix{0.9}{30}{reverse vpath(__v,__cp)}{f8}
 \fmffix{0.5}{30}{reverse vpath(__v,__cp)}{f9}
 \fmffix{0.6}{120}{vpath(__v,__ap)}{f10}
 \fmffix{0.1}{120}{vpath(__v,__ap)}{f11}
 \fmffix{0.9}{30}{reverse vpath(__v,__ap)}{f12}
 \fmffix{0.7}{30}{reverse vpath(__v,__ap)}{f13}
 \fmfipath{num}
 \fmfiset{num}{vloc(__fb)..tension 3..vloc(__fa)..tension 1..vloc(__f0)..tension 3..vloc(__f1)..tension 1..vloc(__f2)..tension 3..vloc(__f3)..tension 1..vloc(__f4)..tension 3..vloc(__f5)..tension 1..vloc(__f6)..tension 3..vloc(__f7)..tension 1..vloc(__f8)..tension 3..vloc(__f9)..tension 1..vloc(__f10)..tension 3..vloc(__f11)..tension 1..vloc(__f12)..tension 3..vloc(__f13)}
 \fmfi{dashes,width=3}{num}
 \fmfcmd{
  shrink 0.7;
  cfill(harrow(num,0.05));
  cfill(harrow(num,0.34));
  cfill(harrow(num,0.60));
  cfill(harrow(num,0.84));
  endshrink;
 }
\end{fmfgraph*}},
\end{equation}
where the arrow indicates the direction of numbering of grassman odd objects and $\cl{\mathcal{D}}$ ($\al{\mathcal{D}}$) stands for $\cd$ ($\ad$)in the case where the leg is connected with another leg by a propagator and for antichiral (chiral) superfield in the case where it is an external leg. The superfields written at each leg are there to remind us with which superfield the leg is associated, they do not necessarily represent superfields attached to the external legs.

Grassman odd objects associated with propagators come from terms in the free field generating functional of the type
\begin{equation*}
 i\int\di^4x\cd\ad\left(\vJ_A\cdot\mathcal{P}\cdot\vJ_B\right),
\end{equation*}
where $\vJ_A$, $\vJ_B$ is an abstract notation for one of the supersources and $\mathcal{P}$ is some operator. The Grassman odd objects in this expression can be numbered from left to right, so the contribution to the supergraph and the impact on the numbering of grassman odd object in the supergraph will look like
\begin{align*}
 &i\times\qquad\parbox{40mm}{\begin{fmfgraph*}(40,10)
  \fmfv{label=$\vphi_A$}{i} \fmfleft{i} \fmfv{label=$\vphi_B$}{o} \fmfright{o} \fmf{plain}{i,o} \fmffreeze
  \fmfopext{0.2}{vpath(__i,__o)}{left}{$\mathcal{D}_A$}{15} \fmfopext{0.35}{vpath(__i,__o)}{left}{$\mathcal{P}$}{20} \fmfopext{0.8}{vpath(__i,__o)}{left}{$\mathcal{D}_B$}{15}
  \fmfv{}{ii,oo} \fmffix{0.9}{100}{reverse vpath(__i,__o)}{ii} \fmffix{0.1}{100}{reverse vpath(__i,__o)}{oo}
  \fmfi{dashes,width=3}{vloc(__ii)--vloc(__oo)} \fmfcmd{shrink 0.7; cfill(harrow(vloc(__ii)--vloc(__oo),0.7));endshrink;}
\end{fmfgraph*}} & &\text{if }\gp{\vphi_A}=0 \text{ and }\gp{\vphi_B}=0, \\
 &i\times\qquad\parbox{40mm}{\begin{fmfgraph*}(40,10)
  \fmfv{label=$\vphi_A$}{i} \fmfleft{i} \fmfv{label=$\vphi_B$}{o} \fmfright{o} \fmf{plain}{i,o} \fmffreeze
  \fmfopext{0.2}{vpath(__i,__o)}{left}{$\mathcal{D}_A$}{15} \fmfopext{0.35}{vpath(__i,__o)}{left}{$\mathcal{P}$}{20}
  \fmfv{}{ii,oo} \fmffix{0.9}{100}{reverse vpath(__i,__o)}{ii} \fmffix{0.1}{100}{reverse vpath(__i,__o)}{oo}
  \fmfi{dashes,width=3}{vloc(__ii)--vloc(__oo)} \fmfcmd{shrink 0.7; cfill(harrow(vloc(__ii)--vloc(__oo),0.7));endshrink;}
\end{fmfgraph*}} & &\text{if }\gp{\vphi_A}=0 \text{ and }\gp{\vphi_B}=1, \\
 &i\times\qquad\parbox{40mm}{\begin{fmfgraph*}(40,10)
  \fmfv{label=$\vphi_A$}{i} \fmfleft{i} \fmfv{label=$\vphi_B$}{o} \fmfright{o} \fmf{plain}{i,o} \fmffreeze
  \fmfopext{0.25}{vpath(__i,__o)}{left}{$\mathcal{P}$}{20} \fmfopext{0.8}{vpath(__i,__o)}{left}{$\mathcal{D}_B$}{15}
  \fmfv{}{ii,oo} \fmffix{0.9}{100}{reverse vpath(__i,__o)}{ii} \fmffix{0.1}{100}{reverse vpath(__i,__o)}{oo}
  \fmfi{dashes,width=3}{vloc(__ii)--vloc(__oo)} \fmfcmd{shrink 0.7; cfill(harrow(vloc(__ii)--vloc(__oo),0.7));endshrink;}
\end{fmfgraph*}} & &\text{if }\gp{\vphi_A}=1 \text{ and }\gp{\vphi_B}=0, \\
 &i\times\qquad\parbox{40mm}{\begin{fmfgraph*}(40,10)
  \fmfv{label=$\vphi_A$}{i} \fmfleft{i} \fmfv{label=$\vphi_B$}{o} \fmfright{o} \fmf{plain}{i,o} \fmffreeze
  \fmfopext{0.25}{vpath(__i,__o)}{left}{$\mathcal{P}$}{20}
  \fmfv{}{ii,oo} \fmffix{0.9}{100}{reverse vpath(__i,__o)}{ii} \fmffix{0.1}{100}{reverse vpath(__i,__o)}{oo}
  \fmfi{dashes,width=3}{vloc(__ii)--vloc(__oo)} \fmfcmd{shrink 0.7; cfill(harrow(vloc(__ii)--vloc(__oo),0.7));endshrink;}
\end{fmfgraph*}} & &\text{if }\gp{\vphi_A}=1 \text{ and }\gp{\vphi_B}=1,
\end{align*}
where $\mathcal{D}_A$ ($\mathcal{D}_B$) is $\ad$ if $\vphi_A$ ($\vphi_B$) is chiral and $\cd$ if it is antichiral.

\subsection{Summary}
With the use of the above results we can summarize the building blocks from which the Feynman supergraphs in our modified Wess-Zumino model are composed.
There is a factor for the supergraph coming from the definition of the effective action
\begin{equation}
 \frac{-i}{(\#\cZ)!(\#\aZ)!(\#\cpsi)!(\#\apsi)!},
\end{equation}
for the superfields attached to the external legs there are factors
\begin{align}
&i\times\qquad\parbox{25mm}{\begin{fmfgraph*}(25,8)
  \fmfv{label=$\cZ$}{i} \fmfleft{i} \fmfv{}{o} \fmfright{o} \fmf{plain}{i,o}
\end{fmfgraph*}}, &
&i\times\qquad\parbox{25mm}{\begin{fmfgraph*}(25,8)
  \fmfv{label=$\aZ$}{i} \fmfleft{i} \fmfv{}{o} \fmfright{o} \fmf{plain}{i,o}
\end{fmfgraph*}}, \\
&i\times\qquad\parbox{25mm}{\begin{fmfgraph*}(25,8)
  \fmfv{label=$\cpsi$}{i} \fmfleft{i} \fmfv{}{o} \fmfright{o} \fmf{plain}{i,o}
\end{fmfgraph*}}, &
&i\times\qquad\parbox{25mm}{\begin{fmfgraph*}(25,8)
  \fmfv{label=$\apsi$}{i} \fmfleft{i} \fmfv{}{o} \fmfright{o} \fmf{plain}{i,o}
\end{fmfgraph*}}, \nonumber
\end{align}
the propagators give expressions
\begin{align}
-i&\times\qquad\parbox{60mm}{\begin{fmfgraph*}(60,14)
 \fmfv{label=$\cZ$}{i} \fmfleft{i}
 \fmfv{label=$\aZ$}{o} \fmfright{o}
 \fmf{plain}{i,o}
 \fmffreeze
 \fmfop{0.35}{vpath(__i,__o)}{left}{$\cfrac{1}{\dbox-m^2-M^2}$}
 \fmfopext{0.1}{vpath(__i,__o)}{left}{$\ad\gi{1}$}{30}
 \fmfopext{0.9}{vpath(__i,__o)}{left}{$\cd\gi{2}$}{30}
 \fmfv{}{a0,a1}
 \fmffix{1}{200}{reverse vpath(__i,__o)}{a0}
 \fmffix{0}{200}{reverse vpath(__i,__o)}{a1}
 \fmfi{dashes,width=3}{vloc(__a0)--vloc(__a1)}
 \fmfcmd{shrink 0.7; cfill(harrow(vloc(__a0)--vloc(__a1),0.6));endshrink;}
\end{fmfgraph*}}, \\
-i&\times\qquad\parbox{60mm}{\begin{fmfgraph*}(60,14)
 \fmfv{label=$\cpsi$}{i} \fmfleft{i}
 \fmfv{label=$\apsi$}{o} \fmfright{o}
 \fmf{plain}{i,o}
 \fmffreeze
 \fmfop{0.3}{vpath(__i,__o)}{left}{$\cfrac{\dbox-m^2}{i\dpp(\dbox-m^2-M^2)}$}
\end{fmfgraph*}}, \nonumber\\
i&\times\qquad\parbox{60mm}{\begin{fmfgraph*}(60,14)
 \fmfv{label=$\cZ$}{i} \fmfleft{i}
 \fmfv{label=$\cpsi$}{o} \fmfright{o}
 \fmf{plain}{i,o}
 \fmffreeze
 \fmfop{0.43}{vpath(__i,__o)}{left}{$\cfrac{\cd\gi{2}}{i\dpp(\dbox-m^2-M^2)}$}
 \fmfopext{0.1}{vpath(__i,__o)}{left}{$\ad\gi{1}$}{30}
 \fmfv{}{a0,a1}
 \fmffix{1}{270}{reverse vpath(__i,__o)}{a0}
 \fmffix{0.4}{270}{reverse vpath(__i,__o)}{a1}
 \fmfi{dashes,width=3}{vloc(__a0)--vloc(__a1)}
 \fmfcmd{shrink 0.7; cfill(harrow(vloc(__a0)--vloc(__a1),0.4));endshrink;}
\end{fmfgraph*}}, \nonumber\\
-i&\times\qquad\parbox{60mm}{\begin{fmfgraph*}(60,14)
 \fmfv{label=$\aZ$}{i} \fmfleft{i}
 \fmfv{label=$\apsi$}{o} \fmfright{o}
 \fmf{plain}{i,o}
 \fmffreeze
 \fmfop{0.43}{vpath(__i,__o)}{left}{$\cfrac{\ad\gi{2}}{i\dpp(\dbox-m^2-M^2)}$}
 \fmfopext{0.1}{vpath(__i,__o)}{left}{$\cd\gi{1}$}{30}
 \fmfv{}{a0,a1}
 \fmffix{1}{270}{reverse vpath(__i,__o)}{a0}
 \fmffix{0.4}{270}{reverse vpath(__i,__o)}{a1}
 \fmfi{dashes,width=3}{vloc(__a0)--vloc(__a1)}
 \fmfcmd{shrink 0.7; cfill(harrow(vloc(__a0)--vloc(__a1),0.4));endshrink;}
\end{fmfgraph*}}, \nonumber
\end{align}
and the vertices are associated with
\begin{align}
i\frac{\lambda}{2}&\times\qquad\parbox{40mm}{\begin{fmfgraph*}(40,40)
 \fmfv{label=$\cZ$}{a} \fmfv{label=$\cZ$}{b} \fmfv{label=$\cpsi$}{c}
 \fmfsurround{d1,d9,d8,a,d2,d3,d4,b,d5,d6,d7,c}
 \fmfv{}{v} \fmf{plain}{a,v} \fmf{plain}{b,v} \fmf{plain}{c,v} \fmfdot{v}
 \fmffreeze
 \fmfop{0.6}{vpath(__v,__c)}{right}{$\cfrac{\cd\gi{1}}{i\dpp}$}
 \fmfopext{0.2}{vpath(__v,__c)}{right}{$\al{\mathcal{D}}\gi{2}$}{20}
 \fmfv{}{a0,a1}
 \fmffix{0.9}{300}{vpath(__v,__c)}{a0}
 \fmffix{0}{300}{vpath(__v,__c)} {a1}
 \fmfi{dashes,width=3}{vloc(__a0)--vloc(__a1)}
 \fmfcmd{shrink 0.7; cfill(harrow(vloc(__a0)--vloc(__a1),0.7));endshrink;}
\end{fmfgraph*}}, &
-i\frac{\lambda}{2}&\times\qquad\parbox{40mm}{\begin{fmfgraph*}(40,40)
 \fmfv{label=$\aZ$}{a} \fmfv{label=$\aZ$}{b} \fmfv{label=$\apsi$}{c}
 \fmfsurround{d1,d9,d8,a,d2,d3,d4,b,d5,d6,d7,c}
 \fmfv{}{v} \fmf{plain}{a,v} \fmf{plain}{b,v} \fmf{plain}{c,v} \fmfdot{v}
 \fmffreeze
 \fmfop{0.6}{vpath(__v,__c)}{right}{$\cfrac{\ad\gi{1}}{i\dpp}$}
 \fmfopext{0.2}{vpath(__v,__c)}{right}{$\cl{\mathcal{D}}\gi{2}$}{20}
 \fmfv{}{a0,a1}
 \fmffix{0.9}{300}{vpath(__v,__c)}{a0}
 \fmffix{0}{300}{vpath(__v,__c)}{a1}
 \fmfi{dashes,width=3}{vloc(__a0)--vloc(__a1)}
 \fmfcmd{shrink 0.7; cfill(harrow(vloc(__a0)--vloc(__a1),0.7));endshrink;}
\end{fmfgraph*}}, \\
i\frac{\lambda}{3}&\times\qquad\parbox{40mm}{\begin{fmfgraph*}(40,40)
 \fmfv{label=$\cZ$}{a} \fmfv{label=$\cZ$}{b} \fmfv{label=$\cZ$}{c}
 \fmfsurround{d1,d9,d8,a,d2,d3,d4,b,d5,d6,d7,c}
 \fmfv{}{v} \fmf{plain}{a,v} \fmf{plain}{b,v} \fmf{plain}{c,v} \fmfdot{v}
 \fmffreeze
 \fmfop{0.5}{vpath(__v,__c)}{right}{$\cfrac{\cd\gi{1}}{i\dpp}$}
 \fmfop{0.5}{vpath(__v,__b)}{left}{$\cfrac{i\dmp\cd\gi{2}}{i\dpp}$}
 \fmfv{}{a0,a1,a2,a3,a4,a5}
 \fmffix{0.9}{300}{vpath(__v,__c)}{a0}
 \fmffix{0.1}{300}{vpath(__v,__c)}{a1}
 \fmffix{0.9}{50}{reverse vpath(__v,__c)}{a2}
 \fmffix{0.1}{50}{reverse vpath(__v,__c)}{a3}
 \fmffix{0.4}{350}{reverse vpath(__v,__b)}{a4}
 \fmffix{1}{350}{reverse vpath(__v,__b)}{a5}
 \fmfipath{arr} 
 \fmfiset{arr}{vloc(__a0)..tension 3..vloc(__a1)..tension 1..vloc(__a2)..tension 3..vloc(__a3)..tension 1..vloc(__a4)..tension 3..vloc(__a5)}
 \fmfi{dashes,width=3}{arr}
 \fmfcmd{shrink 0.7; cfill(harrow(arr,0.2));endshrink;}
 \fmfcmd{shrink 0.7; cfill(harrow(arr,0.85));endshrink;}
\end{fmfgraph*}}, &
-i\frac{\lambda}{3}&\times\qquad\parbox{40mm}{\begin{fmfgraph*}(40,40)
 \fmfv{label=$\aZ$}{a} \fmfv{label=$\aZ$}{b} \fmfv{label=$\aZ$}{c}
 \fmfsurround{d1,d9,d8,a,d2,d3,d4,b,d5,d6,d7,c}
 \fmfv{}{v} \fmf{plain}{a,v} \fmf{plain}{b,v} \fmf{plain}{c,v} \fmfdot{v}
 \fmffreeze
 \fmfop{0.5}{vpath(__v,__c)}{right}{$\cfrac{\ad\gi{1}}{i\dpp}$}
 \fmfop{0.5}{vpath(__v,__b)}{left}{$\cfrac{i\dpm\ad\gi{2}}{i\dpp}$}
 \fmfv{}{a0,a1,a2,a3,a4,a5}
 \fmffix{0.9}{300}{vpath(__v,__c)}{a0}
 \fmffix{0.1}{300}{vpath(__v,__c)}{a1}
 \fmffix{0.9}{50}{reverse vpath(__v,__c)}{a2}
 \fmffix{0.1}{50}{reverse vpath(__v,__c)}{a3}
 \fmffix{0.4}{350}{reverse vpath(__v,__b)}{a4}
 \fmffix{1}{350}{reverse vpath(__v,__b)}{a5}
 \fmfipath{arr} 
 \fmfiset{arr}{vloc(__a0)..tension 3..vloc(__a1)..tension 1..vloc(__a2)..tension 3..vloc(__a3)..tension 1..vloc(__a4)..tension 3..vloc(__a5)}
 \fmfi{dashes,width=3}{arr}
 \fmfcmd{shrink 0.7; cfill(harrow(arr,0.2));endshrink;}
 \fmfcmd{shrink 0.7; cfill(harrow(arr,0.85));endshrink;}
\end{fmfgraph*}} \nonumber.
\end{align}
All the Grassman odd objects (operators and fields) are numbered according to our rules. After we perform the d-algebra on the graph according to the discussion in the next section we compute the final sign by looking at the order of the permutation of the numbers. Finally there are momentum integrals, one for each external field and one for each loop together with a delta function which expresses overall conservation of momentum.

\section{d-algebra}
\label{sec:d}

Supergraph calculations are greatly simplified if one is able manipulate the d-operators involved directly on the supergraph instead of having to write down the mathematical expression and performing the calculations there.
In this section we describe how this can be done in SIM-superspace.

As was mentioned before, if there is an operator in the supergraph we have to distinguish at which end of the propagator it acts. We have a rule which allows us to move derivations from the one end of the propagator to the other. This rule can be schematically expressed as 
\begin{equation}
\label{eq:sgt:move}
 \parbox{30mm}{\begin{fmfgraph*}(30,10)
  \fmfleft{i} \fmfright{o} \fmf{plain}{i,o} \fmfopext{0.2}{vpath(__i,__o)}{left}{$\mathcal{D}$}{15}
 \end{fmfgraph*}}
 \;=\;-\;
 \parbox{30mm}{\begin{fmfgraph*}(30,10)
  \fmfleft{i} \fmfright{o} \fmf{plain}{i,o} \fmfopext{0.8}{vpath(__i,__o)}{left}{$\mathcal{D}$}{15}
 \end{fmfgraph*}},
\end{equation}
where $\mathcal{D}$ is one of $\cd$, $\ad$, $i\pdd{\alpha}{\alpha}$. This means that we have to change the sign before the supergraph each time we move one derivative. The statement is trivial if $\mathcal{D}=i\pdd{\alpha}{\alpha}$, in the case where $\mathcal{D}=\cd,\ad$ it follows from \eqref{eq:ft:ddelta}.

The second thing we can do directly on the supergraph is partial integration, which can be expressed by the rule
\newcommand{\fmfperpartesdiagram}[1]{
 \parbox{30mm}{\begin{fmfgraph*}(30,30)
  \fmfsurroundn{v}{7} \begin{fmffor}{i}{1}{1}{7} \fmf{phantom}{v[i],c} \end{fmffor} \fmfdot{c}
  \fmffreeze
  \begin{fmffor}{i}{1}{2}{5} \fmfi{plain}{vloc(__v[i])--vloc(__c)} \end{fmffor}
  \fmfvn{}{s}{7} \begin{fmffor}{i}{1}{1}{7} \fmffix{0.3}{0}{vpath(__c,__v[i])}{s[i]} \end{fmffor}
  \fmfipath{ang} \fmfiset{ang}{vloc(__s[5])..vloc(__s[6])..vloc(__s[7])..vloc(__s[1])}
  \begin{fmffor}{xang}{0.1}{0.2}{0.9} \fmfiv{decor.shape=circle,decor.filled=full,decor.size=thick}{point xang*length(ang) of ang} \end{fmffor}
  #1
 \end{fmfgraph*}}
}
\begin{equation}
\label{eq:sgt:perpartes}
 \fmfperpartesdiagram{\fmfop{0.5}{vpath(__c,__v[5])}{left}{$\mathcal{D}$}} =
 - \fmfperpartesdiagram{\fmfopext{0.5}{vpath(__c,__v[3])}{right}{$\mathcal{D}$}{10}} 
 - \fmfperpartesdiagram{\fmfopext{0.5}{vpath(__c,__v[1])}{right}{$\mathcal{D}$}{10}} - \cdots,
\end{equation}
where $\mathcal{D}$ is $\cd$ or $\ad$. This rule is a consequence of \eqref{eq:ft:leibniz} \eqref{eq:ft:perpartes}.

Another operation we can perform directly on the supergraph is to apply the $d$-algebra relations \eqref{eq:ft:dalgebra}, which allows us to convert two supercovariant derivatives $\cd$, $\ad$ to one space-time derivative. When we use such relations it is important in which order the covariant derivatives appear in the supergraph, or in another words how are the covariant derivatives numbered as Grassman odd objects. In our case we want to replace two Grassman odd objects by one object, which is not Grassman odd. After they have disappeared, it is not possible to figure out the sign which the supercovariant derivatives gave rise to through anticommutation relations. To remember the way the supercovariant derivatives disappeared we will write the numbers associated with them next to the supergraph.
For example, we have the following identities
\begin{align}
\label{eq:sgt:identity}
 \parbox{25mm}{\begin{fmfgraph*}(25,10)
  \fmfleft{i} \fmfright{o} \fmf{plain}{i,o} \fmffreeze
  \fmfop{0.25}{vpath(__i,__o)}{left}{$\cd\gi{1}\ad\gi{2}\cd\gi{3}$}
 \end{fmfgraph*}} = 
 \parbox{25mm}{\begin{fmfgraph*}(25,10)
  \fmfleft{i} \fmfright{o} \fmf{plain}{i,o} \fmffreeze
  \fmfop{0.25}{vpath(__i,__o)}{left}{$i\dpp\cd\gi{1}$}
 \end{fmfgraph*}}\;2-3 = 
 \parbox{25mm}{\begin{fmfgraph*}(25,10)
  \fmfleft{i} \fmfright{o} \fmf{plain}{i,o} \fmffreeze
  \fmfop{0.25}{vpath(__i,__o)}{left}{$i\dpp\cd\gi{3}$}
 \end{fmfgraph*}}\;1-2, \\
 \parbox{25mm}{\begin{fmfgraph*}(25,10)
  \fmfleft{i} \fmfright{o} \fmf{plain}{i,o} \fmffreeze
  \fmfop{0.25}{vpath(__i,__o)}{left}{$\ad\gi{1}\cd\gi{2}\ad\gi{3}$}
 \end{fmfgraph*}} = 
 \parbox{25mm}{\begin{fmfgraph*}(25,10)
  \fmfleft{i} \fmfright{o} \fmf{plain}{i,o} \fmffreeze
  \fmfop{0.25}{vpath(__i,__o)}{left}{$i\dpp\ad\gi{1}$}
 \end{fmfgraph*}}\;2-3 = 
 \parbox{25mm}{\begin{fmfgraph*}(25,10)
  \fmfleft{i} \fmfright{o} \fmf{plain}{i,o} \fmffreeze
  \fmfop{0.25}{vpath(__i,__o)}{left}{$i\dpp\ad\gi{3}$}
 \end{fmfgraph*}}\;1-2. \nonumber
\end{align}
When we use these identities, we have to be careful about the order of the derivatives, it is important which derivative is closer and which further from the end of the propagator (the notion left and right does not make sense, when the propagator appears in the supergraph). It is also important to distinguish which number is on the left and which on the right side of the dash.

After finishing manipulating the supergraph, we write these pairs at the end of the sequence of numbers which label the grassman odd objects in the expression and if the resulting sequence gives an odd permutation, then we have to add a minus sign.

In the case when the derivatives act on the superfield attached to the external leg we have additional relations following from the fact that the superfield is either chiral or antichiral.
\begin{align}
\label{eq:sgt:external}
 &\parbox{25mm}{\begin{fmfgraph*}(25,10)
  \fmfv{label=$\cl{\vphi}$}{i} \fmfleft{i} \fmfright{o} \fmf{plain}{i,o} \fmffreeze
  \fmfop{0.15}{vpath(__i,__o)}{left}{$\ad$}
 \end{fmfgraph*}} = 0, &
 &\parbox{25mm}{\begin{fmfgraph*}(25,10)
  \fmfv{label=$\cl{\vphi}$}{i} \fmfleft{i} \fmfright{o} \fmf{plain}{i,o} \fmffreeze
  \fmfop{0.2}{vpath(__i,__o)}{left}{$\cd\gi{2}\ad\gi{1}$}
 \end{fmfgraph*}} = \quad
 \parbox{25mm}{\begin{fmfgraph*}(25,10)
  \fmfv{label=$\cl{\vphi}$}{i} \fmfleft{i} \fmfright{o} \fmf{plain}{i,o} \fmffreeze
  \fmfop{0.15}{vpath(__i,__o)}{left}{$i\dpp$}
 \end{fmfgraph*}}\;1-2, \\
 &\parbox{25mm}{\begin{fmfgraph*}(25,10)
  \fmfv{label=$\al{\vphi}$}{i} \fmfleft{i} \fmfright{o} \fmf{plain}{i,o} \fmffreeze
  \fmfop{0.15}{vpath(__i,__o)}{left}{$\cd$}
 \end{fmfgraph*}} = 0, &
 &\parbox{25mm}{\begin{fmfgraph*}(25,10)
  \fmfv{label=$\al{\vphi}$}{i} \fmfleft{i} \fmfright{o} \fmf{plain}{i,o} \fmffreeze
  \fmfop{0.2}{vpath(__i,__o)}{left}{$\ad\gi{2}\cd\gi{1}$}
 \end{fmfgraph*}} = \quad
 \parbox{25mm}{\begin{fmfgraph*}(25,10)
  \fmfv{label=$\al{\vphi}$}{i} \fmfleft{i} \fmfright{o} \fmf{plain}{i,o} \fmffreeze
  \fmfop{0.15}{vpath(__i,__o)}{left}{$i\dpp$}
 \end{fmfgraph*}}\;1-2, \nonumber
\end{align}
where $\cl{\vphi}$ ($\al{\vphi}$) is an arbitrary (anti)chiral superfield.

Finally, to convert the resulting expression into the form where there is only one integral over $\di^2\theta$ we will need the following identity
\begin{equation}
\label{eq:sgt:delta}
 \delta^2(\theta-\theta')\cdot\cd\ad\delta^2(\theta-\theta') = -\delta^2(\theta-\theta')\cdot\ad\cd\delta^2(\theta-\theta') = \delta^2(\theta-\theta').
\end{equation}

Now we are ready to finish our example. The result up until now was given by the  supergraph \eqref{eq:fg:efffirst}. Including the symmetry factor $4$, we use partial integration of the covariant derivatives numbered by $5$ and $6$ which converts this supergraph into
\begin{align*}
&4(-i)\left(i\frac{\lambda}{2}\right)\left(-i\frac{\lambda}{2}\right)(-i^2)i^2\times \\
&\hspace{20mm}\parbox{100mm}{\begin{fmfgraph*}(100,50)
 \fmfv{label=$\cZ$}{i} \fmfleft{i} \fmfv{label=$\aZ$}{o} \fmfright{o}
 \fmfv{}{c} \fmfv{}{a}
 \fmf{plain}{c,i} \fmf{plain}{o,a} \fmf{plain,left,tension=.1,tag=1}{c,a} \fmf{plain,left,tension=.5,tag=2}{a,c}
 \fmffreeze
 \fmfop{0.3}{vpath1(__c,__a)}{left}{$\cfrac{1}{\dbox-m^2-M^2}$}
 \fmfop{0.08}{vpath2(__a,__c)}{left}{$\cd\gi{6}$}
 \fmfop{0.13}{vpath2(__a,__c)}{left}{$\cfrac{\ad\gi{3}}{i\dpp}$}
 \fmfop{0.26}{vpath2(__a,__c)}{left}{$\cd\gi{4}$}
 \fmfop{0.63}{vpath2(__a,__c)}{left}{$\cfrac{\dbox-m^2}{i\dpp(\dbox-m^2-M^2)}$}
 \fmfop{0.74}{vpath2(__a,__c)}{left}{$\ad\gi{2}$}
 \fmfop{0.87}{vpath2(__a,__c)}{left}{$\cfrac{\cd\gi{1}}{i\dpp}$}
 \fmfop{0.93}{vpath2(__a,__c)}{left}{$\ad\gi{5}$}
\end{fmfgraph*}}\qquad.
\end{align*}
We do not get any supergraphs where the covariant derivatives act on the superfields attached to the external legs because of the chirality of the external legs. After using the the identities \eqref{eq:sgt:identity} to reduce the number of covariant derivatives in the loop we obtain the supergraph
\begin{equation*}
-i\lambda^2\times\qquad
\parbox{100mm}{\begin{fmfgraph*}(100,50)
 \fmfv{label=$\cZ$}{i} \fmfleft{i} \fmfv{label=$\aZ$}{o} \fmfright{o}
 \fmfv{}{c} \fmfv{}{a}
 \fmf{plain}{c,i} \fmf{plain}{o,a} \fmf{plain,left,tension=.1,tag=1}{c,a} \fmf{plain,left,tension=.5,tag=2}{a,c}
 \fmffreeze
 \fmfop{0.25}{vpath1(__c,__a)}{left}{$\cfrac{1}{\dbox-m^2-M^2}$}
 \fmfop{0.10}{vpath2(__a,__c)}{left}{$\cd\gi{6}$}
 \fmfop{0.80}{vpath2(__a,__c)}{left}{$\cfrac{\dbox-m^2}{i\dpp(\dbox-m^2-M^2)}$}
 \fmfop{0.90}{vpath2(__a,__c)}{left}{$\ad\gi{5}$}
 \fmfopext{0.20}{vpath2(__a,__c)}{left}{$1-2\quad3-4$}{300}
\end{fmfgraph*}}\qquad.
\end{equation*}
Finally we move the covariant derivative numbered by $5$ to the opposite side of the propagator
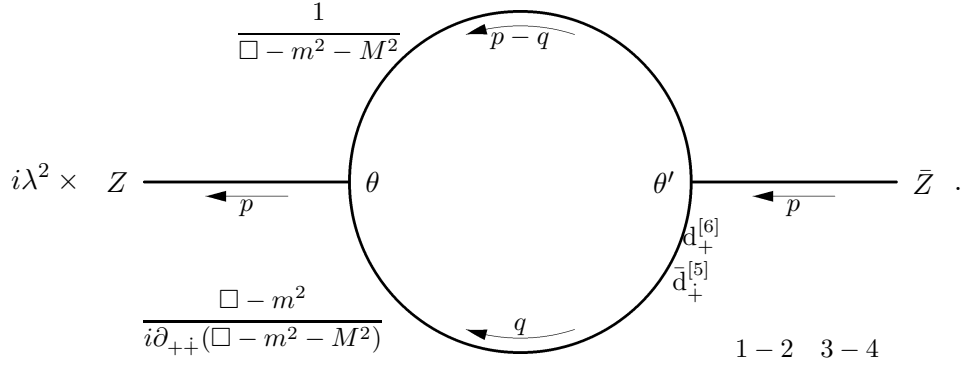
\begin{equation*}
i\lambda^2\times\qquad\parbox{100mm}{\begin{fmfgraph*}(100,50)
 \fmfv{label=$\cZ$}{i} \fmfleft{i}
 \fmfv{label=$\aZ$}{o} \fmfright{o}
 \fmfv{label=$\theta$,label.angle=0}{c}
 \fmfv{label=$\theta'$,label.angle=180}{a}
 \fmf{plain}{c,i} \fmf{plain}{o,a} \fmf{plain,left,tension=.1,tag=1}{c,a} \fmf{plain,left,tension=.5,tag=2}{a,c}
 \fmffreeze
 \fmfop{0.25}{vpath1(__c,__a)}{left}{$\cfrac{1}{\dbox-m^2-M^2}$}
 \fmfop{0.10}{vpath2(__a,__c)}{left}{$\cd\gi{6}$}
 \fmfop{0.15}{vpath2(__a,__c)}{left}{$\ad\gi{5}$}
 \fmfop{0.80}{vpath2(__a,__c)}{left}{$\cfrac{\dbox-m^2}{i\dpp(\dbox-m^2-M^2)}$}
 \fmfopext{0.20}{vpath2(__a,__c)}{left}{$1-2\quad3-4$}{300}
 \fmfarrowpl{0.5}{0.2}{vpath(__c,__i)}{$p$}
 \fmfarrowpl{0.5}{0.2}{vpath(__a,__o)}{$p$}
 \fmfarrownr{0.5}{0.1}{vpath1(__c,__a)}{$p-q$}
 \fmfarrowpr{0.5}{0.1}{vpath2(__c,__a)}{$q$}
\end{fmfgraph*}}\qquad.
\end{equation*}
This is the final supergraph, there are no further simplifications. The expression which corresponds to it is 
\begin{equation*}\begin{split}
 -i\lambda^2
 \int\frac{\di^4p}{(2\pi)^4}\frac{\di^4q}{(2\pi)^4}&\di^2\theta\di^2\theta'\cZ(-p,\theta)\aZ(p,\theta')
 \frac{1}{-(p-q)^2-m^2-M^2} \times \\
 &\frac{-q^2-m^2}{\mpp{q}(-q^2-m^2-M^2)}
 \delta^2(\theta-\theta')\cdot\cd\gi{6}(-q,\theta')\ad\gi{5}(-q'\theta')\delta^2(\theta-\theta').
\end{split}\end{equation*}
The sequence of numbers, which denumber the Grassman odd objects in this expression, together with the pairs of numbers we gathered outside the supergraph gives an odd permutation 
$\left(\begin{smallmatrix}1&2&3&4&5&6\\6&5&1&2&3&4\end{smallmatrix}\right)$
so there was a sign change resulting from the reordering of the Grassman odd objects. We can use the identity \eqref{eq:sgt:delta} to remove the delta function on which the covariant derivatives act, and then remove the remaining delta function by integrating over the $\theta'$ variable. We get the following contribution to the effective action
\begin{equation*}
 i\lambda^2\int\frac{\di^4p}{(2\pi)^4}\di^2\theta\cZ(-p,\theta)\aZ(p,\theta)
 \int\frac{\di^4q}{(2\pi)^4}\frac{q^2+m^2}{\mpp{q}(q^2+m^2+M^2)[(p-q)^2+m^2+M^2]}.
\end{equation*}

\section{A nonrenormalization theorem}
\label{sec:nonren}
It should be clear from the Feynman rules that any supergraph will produce a term which contains a full superspace integral. This means that there can be no direct quantum corrections to chiral terms in the lagrangian, similarly to the case for ordinary Poincare supersymmetric theories. However, just as in the ordinary case, there is a possibility to evade the non renormalization theorem by having nonlocal quantum corrections. In our case we could have a correction
\begin{eqnarray}
\int \di^4 x \cd\ad \cZ^n \frac{\cd}{i\dpp} \cpsi
\end{eqnarray}
which, when pushing in the $\ad$ from the measure, would yield the chiral contribution
\begin{eqnarray}
\int \di^4x \cd \cZ^n \cpsi
\end{eqnarray}
While in the Poincare case, this happens only rarely, in SIM-supersymmetric theories these type of nonlocalities are very common, they are even necessary since SIM-symmetry together with locality implies Lorentz invariance. This means that the nonrenormalization theorem will be much less powerful in the SIM-supersymmetric setting.

\section{Another example}
\label{sec:morex}
For good measure we present a calculation of another supergraph using our Feynman rules and the d-algebra directly on the supergraph. The supergraph will look very much like the supergraph we just calculated, i.e.
\parbox{15mm}{
\begin{fmfgraph*}(15,6)
 \fmfleft{i} \fmfright{o} \fmf{plain}{i,c} \fmf{plain}{a,o} \fmf{plain,left,tension=.5}{c,a,c}
\end{fmfgraph*}},
but this time we will have superfields $\cpsi$, $\apsi$ attached to the external legs. One vertex will be of the type $\cZ\cZ\cpsi$, the other of type $\aZ\aZ\apsi$, which forces us to choose both of the propagators in the loop to be of $\cZ-\aZ$ type. We may compose the supergraph from the building blocks described earlier, which gives us
\begin{align*}
&2(-i)\left(i\frac{\lambda}{2}\right)\left(-i\frac{\lambda}{2}\right)(-i^2)i^2\times \\
&\hspace{20mm}
\parbox{100mm}{\begin{fmfgraph*}(100,43)
 \fmfv{label=$\cpsi\gi{2}$}{i} \fmfleft{i} \fmfv{label=$\apsi\gi{4}$}{o} \fmfright{o}
 \fmfv{}{c} \fmfv{}{a} 
 \fmf{plain}{c,i} \fmf{plain}{o,a} \fmf{plain,left,tension=.1,tag=1}{c,a} \fmf{plain,left,tension=.5,tag=2}{a,c}
 \fmffreeze
 \fmfop{0.13}{vpath1(__c,__a)}{left}{$\ad\gi{5}$}
 \fmfop{0.24}{vpath1(__c,__a)}{left}{$\cfrac{1}{\dbox-m^2-M^2}$}
 \fmfop{0.87}{vpath1(__c,__a)}{left}{$\cd\gi{6}$}
 \fmfop{0.13}{vpath2(__a,__c)}{left}{$\cd\gi{8}$}
 \fmfop{0.79}{vpath2(__a,__c)}{left}{$\cfrac{1}{\dbox-m^2-M^2}$}
 \fmfop{0.87}{vpath2(__a,__c)}{left}{$\ad\gi{7}$}
 \fmfop{0.50}{vpath(__i,__c)}{right}{$\cfrac{\cd\gi{1}}{i\dpp}$}
 \fmfop{0.50}{vpath(__o,__a)}{right}{$\cfrac{\ad\gi{3}}{i\dpp}$}
\end{fmfgraph*}}
\end{align*}
The constant before the supergraph was composed in the following way, the factor $2$ is here because there are two identical graphs contributing, the $-i$ comes from the definition of the effective action, each vertex gives a factor of $\pm\frac{\lambda}{2}$, there are two $-i$, one for each propagator and two $i$, each for one superfield attached to the external leg. The external superfields are Grassman odd, and thus, according to our rules, are numbered in a group with the covariant derivative appearing in the vertex part. The covariant derivative in the loop are numbered together with the propagators.

In the first step we will do an integration by parts with the covariant derivatives numbers $7$ and $8$, which will move them from the loop to the external legs.
\begin{equation*}
-i\frac{\lambda^2}{2}\times\qquad
\parbox{100mm}{\begin{fmfgraph*}(100,45)
 \fmfv{label=$\cpsi\gi{2}$}{i} \fmfleft{i} \fmfv{label=$\apsi\gi{4}$}{o} \fmfright{o}
 \fmfv{}{c} \fmfv{}{a} 
 \fmf{plain}{c,i} \fmf{plain}{o,a} \fmf{plain,left,tension=.1,tag=1}{c,a} \fmf{plain,left,tension=.5,tag=2}{a,c}
 \fmffreeze
 \fmfop{0.13}{vpath1(__c,__a)}{left}{$\ad\gi{5}$}
 \fmfop{0.24}{vpath1(__c,__a)}{left}{$\cfrac{1}{\dbox-m^2-M^2}$}
 \fmfop{0.87}{vpath1(__c,__a)}{left}{$\cd\gi{6}$}
 \fmfop{0.79}{vpath2(__a,__c)}{left}{$\cfrac{1}{\dbox-m^2-M^2}$}
 \fmfop{0.30}{vpath(__i,__c)}{right}{$\ad\gi{7}$}
 \fmfop{0.70}{vpath(__i,__c)}{right}{$\cfrac{\cd\gi{1}}{i\dpp}$}
 \fmfop{0.70}{vpath(__o,__a)}{right}{$\cd\gi{8}$}
 \fmfop{0.30}{vpath(__o,__a)}{right}{$\cfrac{\ad\gi{3}}{i\dpp}$}
\end{fmfgraph*}}
\end{equation*}
In the second step we will use the identities \eqref{eq:sgt:external} to remove operators from the external legs. Moreover we move the covariant derivative number $5$ to the other end of the propagator.
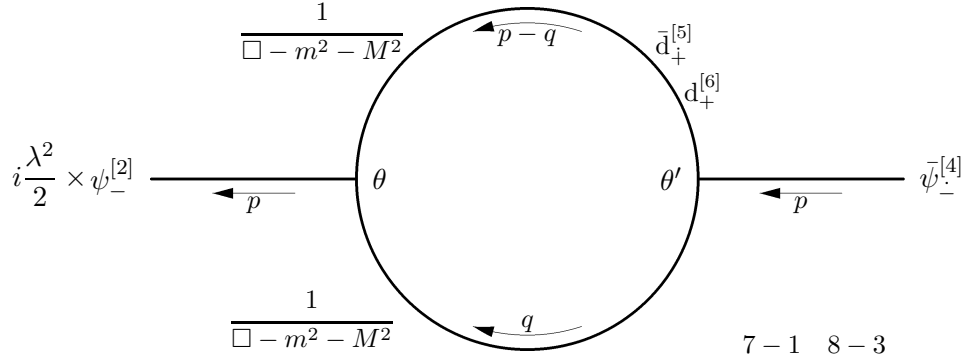
\begin{equation*}
i\frac{\lambda^2}{2}\times\qquad
\parbox{100mm}{\begin{fmfgraph*}(100,45)
 \fmfv{label=$\cpsi\gi{2}$}{i} \fmfleft{i} \fmfv{label=$\apsi\gi{4}$}{o} \fmfright{o}
 \fmfv{label=$\theta$,label.angle=0}{c}
 \fmfv{label=$\theta'$,label.angle=180}{a}
 \fmf{plain}{c,i} \fmf{plain}{o,a} \fmf{plain,left,tension=.1,tag=1}{c,a} \fmf{plain,left,tension=.5,tag=2}{a,c}
 \fmffreeze
 \fmfop{0.24}{vpath1(__c,__a)}{left}{$\cfrac{1}{\dbox-m^2-M^2}$}
 \fmfop{0.77}{vpath1(__c,__a)}{left}{$\ad\gi{5}$}
 \fmfop{0.87}{vpath1(__c,__a)}{left}{$\cd\gi{6}$}
 \fmfop{0.79}{vpath2(__a,__c)}{left}{$\cfrac{1}{\dbox-m^2-M^2}$}
 \fmfopext{0.20}{vpath2(__a,__c)}{left}{$7-1\quad 8-3$}{300}
 \fmfarrowpl{0.5}{0.2}{vpath(__c,__i)}{$p$}
 \fmfarrowpl{0.5}{0.2}{vpath(__a,__o)}{$p$}
 \fmfarrownr{0.5}{0.1}{vpath1(__c,__a)}{$p-q$}
 \fmfarrowpr{0.5}{0.1}{vpath2(__c,__a)}{$q$}
\end{fmfgraph*}}
\end{equation*}
This is the final supergraph. We can immediately remove both delta functions coming from the propagators. The first one is removed by the covariant derivatives $\cd\gi{6}\ad\gi{5}$ according to \eqref{eq:sgt:delta}. Then we can integrate over the second one, which allows us to get rid of the variable $\theta'$. The result is
\begin{equation*}
 -i\frac{\lambda^2}{2}\int\frac{\di^4p}{(2\pi)^4}\di^2\theta\cpsi(-p,\theta)\apsi(p,\theta)
 \int\frac{\di^4q}{(2\pi)^4}\frac{1}{(q^2+m^2+M^2)[(p-q)^2+m^2+M^2]},
\end{equation*}
The sequence of numbers which label the Grassman odd object appearing in this expression together with the sequence $6-5$, labeling the covariant derivatives which we have used to remove one of the delta functions, together with numbers which we gathered outside the supergraph give us the odd permutation
$\left(\begin{smallmatrix}1&2&3&4&5&6&7&8\\2&4&6&5&7&1&8&3\end{smallmatrix}\right)$
so we had to add an extra minus sign.

\section{The effective action in the one loop approximation}
\label{sec:all}

\newcommand{\resdiagd}[6]{
\parbox{24mm}{
\begin{fmfgraph*}(20,15)
 \fmfv{label=\small $#1$,label.dist=1}{i} \fmfleft{i} 
 \fmfv{label=\small $#4$,label.dist=1}{o} \fmfright{o} 
 \fmf{plain}{i,c} \fmf{plain}{a,o} 
 \fmf{plain,left,tension=.3,tag=1}{c,a} 
 \fmf{plain,right,tension=.3,tag=2}{c,a}
 \fmffreeze
 \fmfopext{0.2}{vpath1(__c,__a)}{left}{\scriptsize $#2$}{5}
 \fmfopext{0.2}{vpath2(__c,__a)}{right}{\scriptsize $#3$}{5}
 \fmfopext{0.8}{vpath1(__c,__a)}{left}{\scriptsize $#5$}{5}
 \fmfopext{0.8}{vpath2(__c,__a)}{right}{\scriptsize $#6$}{5}
\end{fmfgraph*}}
}

\newcommand{\resdiagt}[9]{
\parbox{24mm}{
\begin{fmfgraph*}(20,20)
 \fmfsurround{d1,d9,d8,a,d2,d3,d4,b,d5,d6,d7,c}
 \fmfv{label=\small $#1$,label.dist=10}{a}
 \fmfv{label=\small $#7$,label.dist=1}{b}
 \fmfv{label=\small $#4$,label.dist=1}{c}
 \fmf{plain}{a,d} \fmf{plain}{b,e} \fmf{plain}{c,f}
 \fmf{plain,tension=0.2}{d,e} \fmf{plain,tension=0.2}{e,f} \fmf{plain,tension=0.2}{f,d}
 \fmffreeze
 \fmfopext{0.2}{vpath(__d,__e)}{right}{\scriptsize $#2$}{8}
 \fmfopext{0.8}{vpath(__d,__e)}{right}{\scriptsize $#8$}{8}
 \fmfopext{0.2}{vpath(__e,__f)}{right}{\scriptsize $#9$}{12}
 \fmfopext{0.8}{vpath(__e,__f)}{right}{\scriptsize $#6$}{12}
 \fmfopext{0.2}{vpath(__f,__d)}{right}{\scriptsize $#5$}{8}
 \fmfopext{0.8}{vpath(__f,__d)}{right}{\scriptsize $#3$}{8}
\end{fmfgraph*}}
}

In the case at hand, the effective action up to the third order in $\lambda$ and first order in the number of loops can be obtained by summing tree level supergraphs, one loop supergraphs with two external legs and one loop supergraphs with three external legs
The tree level term reproduces the original action, while the one loop contributions are of order $\lambda^2$ for graphs with two external legs and of order $\lambda^3$ for three legged supergraphs. 
By summing these contributions we obtain the effective action valid up to order $\lambda^3$.

As was mentioned the tree level term looks exactly the same as the original action. To compare it with the other terms it is convenient to write it down in the momentum space
\begin{equation*}\begin{split}
 \Gamma_{\text{(tree level)}} &= \int\frac{\di^4p}{(2\pi)^4}\di^2\theta\Big[ 
  \cpsi(-p,\theta)\apsi(p,\theta) - \cZ(-p,\theta)\frac{p^2+m^2}{\mpp{p}}\aZ(p,\theta) \\
  &- iM\cZ(-p,\theta)\frac{\cd(p,\theta)}{\mpp{p}}\cpsi(p,\theta) + iM\aZ(-p,\theta)\frac{\ad(p,\theta)}{\mpp{p}}\apsi(p,\theta)
 \Big] \\
 &+ \int\frac{\di^4p}{(2\pi)^4}\frac{\di^4q}{(2\pi)^4}\di^2\theta\Big[
    i\frac{\lambda}{2}\cZ(p,\theta)\cZ(q,\theta)\frac{\cd(-p-q,\theta)}{\mpp{p}+\mpp{q}}\cpsi(-p-q,\theta) \\
  &- i\frac{\lambda}{3}\cZ(-p-q,\theta)\frac{\cd(p,\theta)}{\mpp{p}}\cZ(p,\theta)\frac{\mmp{q}\cd(q,\theta)}{\mpp{q}}\cZ(q,\theta) \\
  &- i\frac{\lambda}{2}\aZ(p,\theta)\aZ(q,\theta)\frac{\ad(-p-q,\theta)}{\mpp{p}+\mpp{q}}\apsi(-p-q,\theta) \\
  &+ i\frac{\lambda}{3}\aZ(-p-q,\theta)\frac{\ad(p,\theta)}{\mpp{p}}\aZ(p,\theta)\frac{\mpm{q}\ad(q,\theta)}{\mpp{q}}\aZ(q,\theta)
 \Big].
\end{split}\end{equation*}

The contribution from one loop supergraphs with two legs comes from the supergraphs summarized in the following pictures
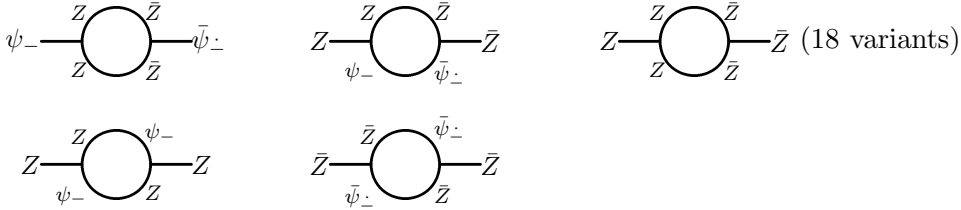
\begin{align*}
 &\resdiagd{\cpsi}{\cZ}{\cZ}{\apsi}{\aZ}{\aZ} &
 &\resdiagd{\cZ}{\cZ}{\cpsi}{\aZ}{\aZ}{\apsi} &
 &\resdiagd{\cZ}{\cZ}{\cZ}{\aZ}{\aZ}{\aZ}\text{(18 variants)} \\
 &\resdiagd{\cZ}{\cZ}{\cpsi}{\cZ}{\cpsi}{\cZ} &
 &\resdiagd{\aZ}{\aZ}{\apsi}{\aZ}{\apsi}{\aZ}
\end{align*}
There are $18$ supergraphs corresponding to the third picture, because the vertices of type $\cZ\cZ\cZ$ and $\aZ\aZ\aZ$ are not symmetric with respect to permutation of the external legs.
The supergraphs depicted in the last two pictures are equal to zero, because the resulting expressions has the integral $\int\di^2\theta$ over the product of two superfields, both of them being either chiral or antichiral.
The overall result can be written as
\begin{equation*}\begin{split}
 &\Gamma_{\text{(one loop 2 legs)}} = -i\frac{\lambda^2}{2}\int\frac{\di^4p}{(2\pi)^4}\di^2\theta\left(\cpsi(-p,\theta)\apsi(p,\theta)-\cZ(-p,\theta)\frac{p^2}{\mpp{p}}\aZ(p,\theta)\right)\times \\
 &\qquad\qquad\qquad\qquad\qquad\qquad\qquad\int\frac{\di^4q}{(2\pi)^4}\frac{1}{(q^2+m^2+M^2)[(p-q)^2+m^2+M^2]} \\
 &\qquad +i\lambda^2\int\frac{\di^4p}{(2\pi)^4}\di^2\theta\cZ(-p,\theta)\aZ(p,\theta)\int\frac{\di^4q}{(2\pi)^4}\frac{m^2}{\mpp{q}(q^2+m^2+M^2)[(p-q)^2+m^2+M^2]}.
\end{split}\end{equation*}
The ultraviolet convergence of the integrals appearing in the expressions can be investigated by means of power counting. The first integral is logarithmically divergent, while the second integral is convergent in four dimensions.
We can calculate the integrals with the help of dimensional regularization in $4 - 2\epsilon$ dimensions
\begin{equation*}\begin{split}
 &\Gamma_{\text{(one loop 2 legs)}} = \frac{\lambda^2}{2(4\pi)^2}\int\frac{\di^4p}{(2\pi)^4}\di^2\theta\Bigg[ \\
 &\quad\left(\cpsi(-p,\theta)\apsi(p,\theta)-\cZ(-p,\theta)\frac{p^2}{\mpp{p}}\aZ(p,\theta)\right)\left(\frac{1}{\epsilon} + I_1(p^2)\right) -\frac{m^2}{\mpp{p}}\cZ(-p,\theta)\aZ(p,\theta)I_2(p^2)\Bigg].
\end{split}\end{equation*}
where the functions $I_1$ and $I_2$ are
\begin{align*}
 I_1(p^2) &= -\gamma + \ln\frac{4\pi e^2}{m^2+M^2}
  - \sqrt{1+4\frac{m^2+M^2}{p^2}}\ln\left(\frac{\sqrt{1+4\frac{m^2+M^2}{p^2}}+1}{\sqrt{1+4\frac{m^2+M^2}{p^2}}-1}\right), \\
 I_2(p^2) &= -\ln^2\left(\frac{\sqrt{1+4\frac{m^2+M^2}{p^2}}+1}{\sqrt{1+4\frac{m^2+M^2}{p^2}}-1}\right).
\end{align*}
Because one of the integrals was logarithmically divergent, there is a factor $\epsilon^{-1}$ which is divergent in the limit where the dimension approaches $4$, i.e. when $\epsilon\rightarrow 0$.

The contribution from the one loop supergraphs with three legs comes from the supergraphs summarized in the following pictures
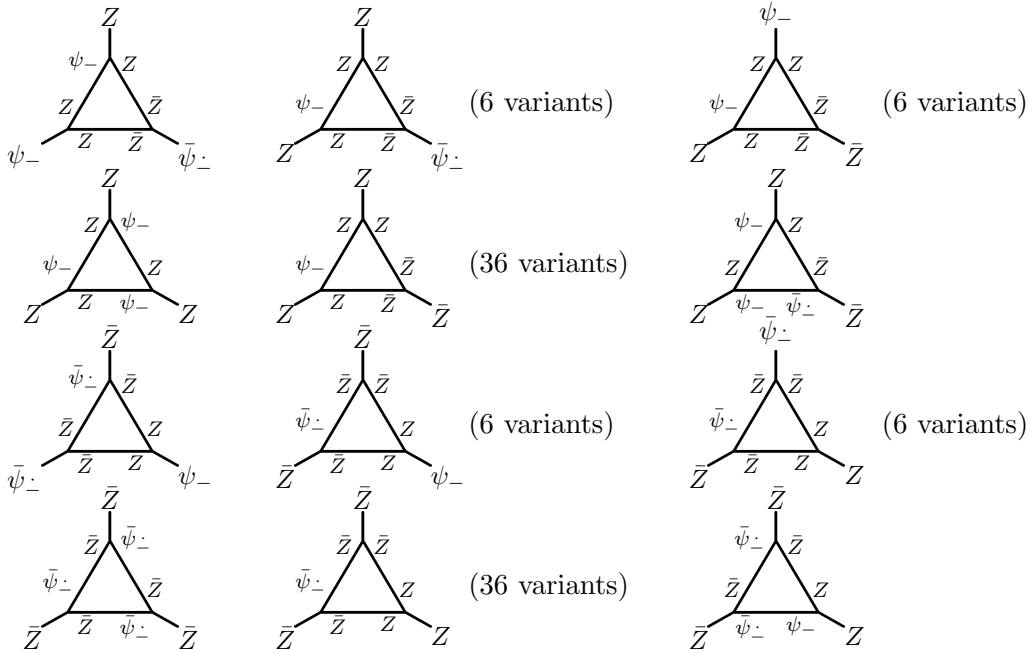
\begin{align*}
 &\resdiagt{\cZ}{\cpsi}{\cZ}{\apsi}{\aZ}{\aZ}{\cpsi}{\cZ}{\cZ} &
 &\resdiagt{\cZ}{\cZ}{\cZ}{\apsi}{\aZ}{\aZ}{\cZ}{\cpsi}{\cZ}\text{(6 variants)} &
 &\resdiagt{\cpsi}{\cZ}{\cZ}{\aZ}{\aZ}{\aZ}{\cZ}{\cpsi}{\cZ}\text{(6 variants)} \\
 &\resdiagt{\cZ}{\cZ}{\cpsi}{\cZ}{\cZ}{\cpsi}{\cZ}{\cpsi}{\cZ} &
 &\resdiagt{\cZ}{\cZ}{\cZ}{\aZ}{\aZ}{\aZ}{\cZ}{\cpsi}{\cZ}\text{(36 variants)} &
 &\resdiagt{\cZ}{\cpsi}{\cZ}{\aZ}{\aZ}{\apsi}{\cZ}{\cZ}{\cpsi} \\
 &\resdiagt{\aZ}{\apsi}{\aZ}{\cpsi}{\cZ}{\cZ}{\apsi}{\aZ}{\aZ} &
 &\resdiagt{\aZ}{\aZ}{\aZ}{\cpsi}{\cZ}{\cZ}{\aZ}{\apsi}{\aZ}\text{(6 variants)} &
 &\resdiagt{\apsi}{\aZ}{\aZ}{\cZ}{\cZ}{\cZ}{\aZ}{\apsi}{\aZ}\text{(6 variants)} \\
 &\resdiagt{\aZ}{\aZ}{\apsi}{\aZ}{\aZ}{\apsi}{\aZ}{\apsi}{\aZ} &
 &\resdiagt{\aZ}{\aZ}{\aZ}{\cZ}{\cZ}{\cZ}{\aZ}{\apsi}{\aZ}\text{(36 variants)} &
 &\resdiagt{\aZ}{\apsi}{\aZ}{\cZ}{\cZ}{\cpsi}{\aZ}{\aZ}{\apsi}
\end{align*}
The supergraphs depicted in the last six pictures can be obtained from the supergraphs depicted in the first six pictures by complex conjugation. The supergraphs depicted in the third, fourth, ninth and tenth picture are equal to zero, the reason is again that the expression would be a full superspace integral of something chiral. The expressions for the supergraph depicted in the fourth is proportional to the integral $\int\di^2\theta$ over the chiral product of superfields $\cZ\cZ\cZ$, while the expressions for supergraphs depicted in the third picture contains the integral $\int\di^2\theta$ over the chiral product $(\ad\aZ)\cZ\cpsi$ while the tenth and ninth graphs are given by the complex conjugate expressions. The overall expression for the sum of all the diagrams is
\begin{equation*}\begin{split}
 &\Gamma_{\text{(one loop 3 legs)}} = \frac{\lambda^3}{2}\int\frac{\di^4p}{(2\pi)^4}\frac{\di^4q}{(2\pi)^4}\di^2\theta\Bigg[ 2\cpsi(p,\theta)\apsi(q,\theta)\cZ(-p-q,\theta) \\
 &\qquad +2\frac{\mmp{p}}{\mpp{p}}\cd(p,\theta)\cZ(p,\theta)\cZ(q,\theta)\apsi(-p-q,\theta) +\cZ(p,\theta)\cZ(q,\theta)\frac{(p+q)^2}{\mpp{p}+\mpp{q}}\aZ(-p-q,\theta) \\
 &\qquad- 2\apsi(p,\theta)\cpsi(q,\theta)\aZ(-p-q,\theta) -2\frac{\mpm{p}}{\mpp{p}}\ad(p,\theta)\aZ(p,\theta)\aZ(q,\theta)\cpsi(-p-q,\theta) \\
 &\qquad-\aZ(p,\theta)\aZ(q,\theta)\frac{(p+q)^2}{\mpp{p}+\mpp{q}}\cZ(-p-q,\theta) \Bigg] \times \\
 &\qquad \int\frac{\di^4r}{(2\pi)^4}\frac{1}{(r^2+m^2+M^2)[(p-r)^2+m^2+M^2][(q+r)^2+m^2+M^2]} \\
 &\qquad +\lambda^3\int\frac{\di^4p}{(2\pi)^4}\frac{\di^4q}{(2\pi)^4}\di^2\theta\Bigg[ -\cZ(p,\theta)\cZ(q,\theta)\aZ(-p-q,\theta) \\
 &\qquad +\aZ(p,\theta)\aZ(q,\theta)\cZ(-p-q,\theta) \Bigg] \times \\
 &\qquad \int\frac{\di^4r}{(2\pi)^4}\frac{m^2}{(\mpp{r}-\mpp{p})(r^2+m^2+M^2)[(p-r)^2+m^2+M^2][(q+r)^2+m^2+M^2]}.
\end{split}\end{equation*}
As in the case of the two legged one loop supergraphs, the ultraviolet convergence of the integrals can be investigated by power counting. Both of the integrals appearing in the expression are convergent.

\newcommand{\rZ}{\mathcal{Z}}

\section{Renormalization}
\label{sec:ren}

This section is devoted to the renormalization of our model using the minimal subtraction scheme in the one loop approximation. The asymptotic behavior will be investigated at the end of the section.

To renormalize the effective action, we have to replace the bare superfields $\cZ,\aZ,\cpsi,\apsi$ with their renormalized counterparts $\cZ_R,\aZ_R,\cpsi{}_R,\apsi{}_R$. It is enough to have one renormalization factor for superfields $\cZ,\aZ$ and one for superfields $\cpsi,\apsi$, because the superfields in each pair are related by complex conjugation and it is reasonable to expect that the renormalization factor is the same
\begin{align*}
 \cZ &= \sqrt{\rZ_1}\cZ_R, &
 \aZ &= \sqrt{\rZ_1}\aZ_R, &
 \cpsi &= \sqrt{\rZ_2}\cpsi{}_R, &
 \apsi &= \sqrt{\rZ_2}\apsi{}_R.
\end{align*}

In the minimal subtraction scheme we have the following expressions for the bare parameters and wave function renormalization factors
\begin{align}
 \lambda &= \mu^\epsilon\left(\lambda_R + \sum_{n=1}^{\infty}\frac{\lambda_{(n)}}{\epsilon^n}\right), & 
 m^2 &= m^2_R\left(1 + \sum_{n=1}^{\infty}\frac{m^2_{(n)}}{\epsilon^n}\right), &
 M &= M_R\left(1 + \sum_{n=1}^{\infty}\frac{M_{(n)}}{\epsilon^n}\right), \nonumber \\
 \rZ_1 &= 1 + \sum_{n=1}^{\infty}\frac{\rZ_{1(n)}}{\epsilon^n}, &
 \rZ_2 &= 1 + \sum_{n=1}^{\infty}\frac{\rZ_{2(n)}}{\epsilon^n},
 \label{eq:rg:bare}
\end{align}
where $\mu$ is a parameter with dimension of mass. The parameters $\lambda_{(n)}$, $m^2_{(n)}$, $M_{(n)}$, $\rZ_{1(n)}$ and $\rZ_{2(n)}$ have to be chosen in such a way, that the effective action expressed in the renormalized quantities does not have any terms with negative powers of $\epsilon$. They may depend only on $\lambda_R$, $m^2_R$ and $M_R$ (we will see later, that they actually depend only on $\lambda_R$). The factor $\mu^\epsilon$ is included in the definition of $\lambda$ to compensate for the change in its dimension when the space-time dimension is changed, so the renormalized parameter $\lambda_R$ is always dimensionless.

If we look at the factors in front of the different terms in the effective action, we see, that the effective action will not contain $\epsilon$ poles if the following expressions are free of $\epsilon$ poles.
\begin{align*}
 &\rZ_1\left(1+\frac{1}{\epsilon}\frac{\lambda^2}{2(4\pi)^2}\right), &
 &\rZ_2\left(1+\frac{1}{\epsilon}\frac{\lambda^2}{2(4\pi)^2}\right), &
 &\rZ_1m^2, \\
 &\sqrt{\rZ_1\rZ_2}M, &
 &\sqrt{\rZ_1^2\rZ_2}\lambda, &
 &\sqrt{\rZ_1^3}\lambda.
\end{align*}
After substituting \eqref{eq:rg:bare}, we find out that we have to choose
\begin{align*}
 \lambda &= \mu^\epsilon\left[\lambda_R + \frac{1}{\epsilon}\left(\frac{3\lambda_R^3}{4(4\pi)^2} + \mathcal{O}(\lambda_R^4)\right)\right], &
 m^2 &= m^2_R\left[1 + \frac{1}{\epsilon}\left(\frac{\lambda_R^2}{2(4\pi)^2} + \mathcal{O}(\lambda_R^3)\right)\right], \\
 M &= M_R\left[1 + \frac{1}{\epsilon}\left(\frac{\lambda_R^2}{2(4\pi)^2} + \mathcal{O}(\lambda_R^3)\right)\right], &
 \rZ_1 &= 1 - \frac{1}{\epsilon}\left(\frac{\lambda_R^2}{2(4\pi)^2} + \mathcal{O}(\lambda_R^3)\right), \\
 \rZ_2 &= 1 - \frac{1}{\epsilon}\left(\frac{\lambda_R^2}{2(4\pi)^2} + \mathcal{O}(\lambda_R^3)\right),
\end{align*}

We are interested in the dependence of the renormalized parameters on the renormalization scale, i.e. the dependence on the parameter $\mu$. This dependence is fully described by the functions
\begin{align}
 \beta &= \mu\frac{\di}{\di\mu}\lambda_R, &
 \gamma_{m^2} &= \frac{1}{m_R^2}\mu\frac{\di}{\di\mu}m_R^2, &
 \gamma_{M} &= \frac{1}{M_R}\mu\frac{\di}{\di\mu}M_R, \nonumber \\
 \gamma_{\rZ_1} &= \frac{1}{\rZ_1}\mu\frac{\di}{\di\mu}\rZ_1, &
 \gamma_{\rZ_2} &= \frac{1}{\rZ_2}\mu\frac{\di}{\di\mu}\rZ_2.
\label{eq:rg:betagamma}
\end{align}
To obtain these functions, it is enough to know the divergent terms coming from the corrections to the effective action. We know that the bare parameters do not depend on the scale $\mu$. This condition can be expressed by the following set of equations
\begin{align*}
 0 &= \frac{1}{\mu^\epsilon}\mu\frac{\di}{\di\mu}\lambda = \epsilon\left(\lambda_R + \frac{1}{\epsilon}\frac{3\lambda_R^3}{4(4\pi)^2}\right) 
  + \beta\left(1 + \frac{1}{\epsilon}\frac{\di}{\di\lambda_R}\frac{3\lambda_R^3}{4(4\pi)^2}\right) + \mathcal{O}(\lambda_R^4), \\
 0 &= \frac{1}{m^2_R}\mu\frac{\di}{\di\mu}m^2 = \gamma_{m^2}\left(1 + \frac{1}{\epsilon}\frac{\lambda_R^2}{2(4\pi)^2}\right)
  + \frac{1}{\epsilon}\beta\frac{\di}{\di\lambda_R}\frac{\lambda_R^2}{2(4\pi)^2} + \mathcal{O}(\lambda_R^3), \\
 0 &= \frac{1}{M_R}\mu\frac{\di}{\di\mu}M = \gamma_{M}\left(1 + \frac{1}{\epsilon}\frac{\lambda_R^2}{2(4\pi)^2}\right)
  + \frac{1}{\epsilon}\beta\frac{\di}{\di\lambda_R}\frac{\lambda_R^2}{2(4\pi)^2} + \mathcal{O}(\lambda_R^3).
\end{align*}
Similar equations can be constructed also for the pair of wave function renormalization $\gamma$ functions
\begin{align*}
 \mu\frac{\di}{\di\mu}\rZ_1 &= \left(1 - \frac{1}{\epsilon}\frac{\lambda_R^2}{2(4\pi)^2}\right)\gamma_{\rZ_1}
  = -\frac{1}{\epsilon}\beta\frac{\di}{\di\lambda_R}\frac{\lambda_R^2}{2(4\pi)^2} + \mathcal{O}(\lambda_R^3), \\
 \mu\frac{\di}{\di\mu}\rZ_2 &= \left(1 - \frac{1}{\epsilon}\frac{\lambda_R^2}{2(4\pi)^2}\right)\gamma_{\rZ_2}
  = -\frac{1}{\epsilon}\beta\frac{\di}{\di\lambda_R}\frac{\lambda_R^2}{2(4\pi)^2} + \mathcal{O}(\lambda_R^3).
\end{align*}
The $\beta$ and $\gamma$ functions in these equations can be expanded in powers of $\epsilon$
\begin{align*}
 \beta &= \sum_{n=0}^{\infty}\beta_{(n)}\epsilon^n, &
 \gamma_{m^2} &= \sum_{n=0}^{\infty}\gamma_{m^2(n)}\epsilon^n, &
 \gamma_{M} &= \sum_{n=0}^{\infty}\gamma_{M(n)}\epsilon^n, \\
 \gamma_{\rZ_1} &= \sum_{n=0}^{\infty}\gamma_{\rZ_1(n)}\epsilon^n, &
 \gamma_{\rZ_2} &= \sum_{n=0}^{\infty}\gamma_{\rZ_2(n)}\epsilon^n,
\end{align*}
then we require that the equations holds independently for each power of $\epsilon$. This gives us a set of equations, from which we can obtain the coefficients in the expansion of the $\beta$ and $\gamma$ functions. In the limit where $\epsilon\rightarrow 0$ we obtain
\begin{align*}
 \beta &= \frac{3}{2}\frac{\lambda_R^3}{(4\pi)^2}+\mathcal{O}(\lambda_R^4), &
 \gamma_{m^2} &= \frac{\lambda_R^2}{(4\pi)^2}+\mathcal{O}(\lambda_R^3), &
 \gamma_{M} &= \frac{\lambda_R^2}{(4\pi)^2}+\mathcal{O}(\lambda_R^3), \\
 \gamma_{\rZ_1} &= \frac{\lambda_R^2}{(4\pi)^2}+\mathcal{O}(\lambda_R^3), &
 \gamma_{\rZ_2} &= \frac{\lambda_R^2}{(4\pi)^2}+\mathcal{O}(\lambda_R^3).
\end{align*}

The dependence of the renormalized parameters and the wave function renormalization factors on the scale $\mu$ in the one loop approximation can be obtained by solving the differential equations \eqref{eq:rg:betagamma}. If $\mu'$ is some renormalization scale, at which we know the values of renormalized parameters $\lambda_R(\mu'),m_R^2(\mu'),M_R(\mu')$ and the wave function renormalization factors $\rZ_1(\mu'),\rZ_2(\mu')$, then the renormalized parameters $\lambda_R(\mu),m_R^2(\mu),M_R(\mu)$ and wave function renormalization factors $\rZ_1(\mu),\rZ_2(\mu)$ at some other renormalization scale $\mu$ are equal to
\begin{align*}
 \lambda_R(\mu) &= \lambda_R(\mu')\left(1-\frac{3[\lambda_R(\mu')]^2}{(4\pi)^2}\ln\frac{\mu}{\mu'}\right)^{-\frac{1}{2}}, &
 m^2_R(\mu) &= m_R^2(\mu')\left(1-\frac{3[\lambda_R(\mu')]^2}{(4\pi)^2}\ln\frac{\mu}{\mu'}\right)^{-\frac{1}{3}}, \\
 M_R(\mu) &= M_R(\mu')\left(1-\frac{3[\lambda_R(\mu')]^2}{(4\pi)^2}\ln\frac{\mu}{\mu'}\right)^{-\frac{1}{3}}, &
 \rZ_1(\mu) &= \rZ_1(\mu')\left(1-\frac{3[\lambda_R(\mu')]^2}{(4\pi)^2}\ln\frac{\mu}{\mu'}\right)^{-\frac{1}{3}}, \\
 \rZ_2(\mu) &= \rZ_2(\mu')\left(1-\frac{3[\lambda_R(\mu')]^2}{(4\pi)^2}\ln\frac{\mu}{\mu'}\right)^{-\frac{1}{3}}.
\end{align*}
From these equations it can be seen that when the renormalization scale approaches zero all of the renormalized parameters $\lambda_R,m^2_R,M_R$ and wave function renormalization factors $\rZ_1,\rZ_2$ approaches zero too. On the other hand if the renormalization scale approaches the value $\mu\rightarrow\mu'\exp\left(\frac{(4\pi)^2}{3[\lambda(\mu')]^2}\right)$ then all parameters $\lambda_R,m^2_R,M_R,\rZ_1,\rZ_2$ diverge to infinity.
It is also interesting to note that the two different masses $(m,M)$ renormalize differently, in fact, $\frac{m_R^2}{M_R}$ is invariant under renormalization.

\section{Conclusions}
\label{sec:concl}
In this paper we constructed Feynman rules for the SIM(2) superspace formulation \cite{uam} of the Wess-Zumino model, including a term which manifestly breaks Lorentz invariance. To check the consistency of our formalism, we performed a one loop calculation of the effective action to see that the results agree with known results for the Wess-Zumino model when the Lorentz violating mass term is put to zero. We calculated the renormalization of all masses and coupling constants of the theory. We found that the presence of terms invariant under only the SIM(2) supergroup did not affect the properties of the model destructively. Interestingly, we found that the supersymmetric mass $M$ and the SIM-symmetric mass $m$ renormalize differently. In particular, $m$ renormalize to zero more slowly than $M$ at low energies.

There are several interesting open questions to which we would like to return in the future. Using our results it would be possible to start speculating about the role of SIM-supersymmetry in the Supersymmetric Standard Model. However, to do that it would be desirable to first address the question about SIM-supersymmetric gauge multiplets and its quantization. Other intersting questions that we would like to address is the target space geometry of SIM-supersymmetric sigma models.

\acknowledgments
Discussions with Ulf Lindstr\"om and Martin Ro\v{c}ek are gratefully acknowledged. The research of S.P and J.V. was supported by the Czech government grant agency under contract no. GA\v{C}R 202/08/H072. The work of RvU was supported by a Czech Ministry of Education grant No. MSM0021622409. 

\end{fmffile}
\end{document}